\documentclass[conference]{IEEEtran}

\IEEEoverridecommandlockouts

\usepackage[breaklinks=true]{hyperref}
\hypersetup{colorlinks,citecolor=blue,linkcolor = blue}

\usepackage{tikz}
\usetikzlibrary{arrows,shapes}
\usepackage{dot2texi}

\usepackage{enumitem} 
\usepackage{amsfonts}
\usepackage[english]{babel}
\usepackage[utf8]{inputenc}
\usepackage{graphicx}
\usepackage{amssymb,amsmath,amsthm}
\usepackage{mathrsfs}
\usepackage{mathtools}
\usepackage{bm}
\usepackage{adjustbox}   
\usepackage{subfig}

\usepackage{tabularx,colortbl}
\usepackage{colortbl}
\usepackage{graphicx}
\usepackage{listings}
\usepackage{cleveref}
\usepackage{soul}
\usepackage{url}
\usepackage{setspace}
\usepackage{makecell}
\usepackage{algorithm}
\usepackage{algorithmic}

\def\BibTeX{{\rm B\kern-.05em{\sc i\kern-.025em b}\kern-.08emT\kern-.1667em\lower.7ex\hbox{E}\kern-.125emX}}

\date{\today}

\newcommand{\R}{\mathbb{R}}

\newcommand{\tsr}[1]{\pmb{\mathcal{#1}}}
\newcommand{\proc}[1]{\mathcal{#1}}

\newcommand{\vcr}[1]{\mathbf{#1}}
\newcommand{\mat}[1]{\mathbf{#1}}
\newcommand{\mb}[2]{\mathbf{#1}^{(#2)}}
\newcommand{\mt}[3]{\mathbf{#1}^{(#2)}_{#3}}
\newcommand{\defeq}{\coloneqq}

\newcommand{\fnrm}[1]{{\lvert \lvert #1 \rvert \rvert}_F}

\newcommand{\vnrm}[1]{{\lvert \lvert #1 \rvert \rvert}_2}
\newcommand{\inti}[2]{\{{#1},\ldots, {#2}\}}
\newcommand{\M}{M}

\newcommand{\name}[1]{{\color{blue}[name] }}

\newcommand {\Comment}[1]{{\small{\textit{//   #1}}}}
\newcommand{\bigast}{\mathop{\scalebox{2.}{\raisebox{-0.2ex}{$\ast$}}}}%
\newcommand{\bigo}{\mathcal{O}}

\usepackage[normalem]{ulem}

\usepackage{pythonhighlight}

\usepackage{listings}
\definecolor{mygreen}{rgb}{0,0.2,0}
\definecolor{mygray}{rgb}{0.5,0.5,0.5}
\definecolor{mymauve}{rgb}{0.58,0,0.82}
\definecolor{mypurple}{rgb}{0.38,0,0.32}
\definecolor{myblue}{rgb}{0.1,0,0.32}
\newcommand{\costyle}{\footnotesize\ttfamily\bfseries}
\newcommand{\kwstyle}{\costyle\textcolor{myblue}}
\begin{document}
\pagestyle{plain}

\title{Efficient parallel CP decomposition with pairwise perturbation and multi-sweep dimension tree}

\author{
\IEEEauthorblockN{
    Linjian Ma
    and
    Edgar Solomonik
}
\IEEEauthorblockA{
    Department of Computer Science, 
    University of Illinois at Urbana-Champaign \\
    Email:
    \{lma16, solomon2\}@illinois.edu
}
}

\maketitle

\begin{abstract}
CP tensor decomposition with alternating least squares (ALS) is dominated in cost by the matricized-tensor
times Khatri-Rao product (MTTKRP) kernel that is necessary to set up the quadratic optimization subproblems. State-of-art parallel ALS implementations use dimension
trees to avoid redundant computations across MTTKRPs within
each ALS sweep.
In this paper, we propose two new parallel algorithms to accelerate CP-ALS. 
We introduce the multi-sweep dimension tree (MSDT) algorithm, which requires the contraction between an order $N$ input tensor and the first-contracted input matrix once every $(N-1)/N$ sweeps.
This algorithm reduces the leading order computational cost by a factor of $2(N-1)/N$ relative to the best previously known approach.
In addition, we introduce a 
more communication-efficient approach to parallelizing an approximate CP-ALS algorithm, pairwise perturbation.
This technique uses perturbative corrections to the subproblems rather than recomputing the contractions, and asymptotically accelerates ALS. 
Our benchmark results show that the per-sweep time achieves 1.25X speed-up for MSDT and 1.94X speed-up for pairwise perturbation compared to the state-of-art dimension trees running on 1024 processors on the Stampede2 supercomputer. 
\end{abstract}

\begin{IEEEkeywords}
CP decomposition, dimension tree, pairwise perturbation
\end{IEEEkeywords}

\section{Introduction}
\label{sec:intro}
The CANDECOMP/PARAFAC (CP) decomposition expresses the input tensor as a sum of rank-one tensors, and is widely used in signal processing~\cite{murphy2013fluorescence,sidiropoulos2017tensor,maruhashi2011multiaspectforensics,cong2015tensor}, quantum chemistry~\cite{hohenstein2012tensor,hummel2017low} and machine learning applications~\cite{anandkumar2014tensor,kolda2009tensor}. 
CP tensor decomposition can be computed via different optimization techniques, such as variants of gradient descent and Newton's method~\cite{acar2011scalable,phan2013low,paatero1997weighted,singh2019comparison}
and alternating least squares~\cite{carroll1970analysis,harshman1970foundations}. 
Of them, the alternating least squares (ALS) method is most commonly used as it makes  relatively large amount of progress and is computationally cheap for each iteration, and guarantees the monotonic decrease of the decomposition residual.

The CP-ALS procedure consists of \textit{sweeps}, and each sweep updates every factor matrix once in a fixed order.
The procedure for updating one factor matrix can be formulated as the optimization subproblem that only regards the target factor matrix as the optimization variable. Each optimization subproblem is quadratic, which guarantees the decrease of the decomposition residual.

In each sweep of CP-ALS, the computational cost is dominated by the operations necessary for the right-hand-side constructions for each normal equation in the optimization subproblem. This operation, called \textit{the matricized tensor-times Khatri-Rao product} (MTTKRP), involves multiple \textit{tensor-times-matrix} (TTM) and \textit{tensor-times-vector} (TTV) operations when the input tensor is in a dense format. 
For an order $N$ tensor with modes of dimension $s$ and CP rank $R$, $N$ MTTKRPs are necessary in each sweep, each costing $2s^NR$ to the leading order.
The state-of-art dimension tree based construction of MTTKRP~\cite{kaya2019computing,phan2013fast,ballard2018parallel} uses amortization to save the cost, and has been implemented in multiple tensor computation libraries~\cite{eswar2019planc,ma2020autohoot}. However, it still requires the computational cost of at least $4s^{N}R$ for each sweep.
To further
accelerate CP-ALS, many researchers leverage different techniques to \textit{parallelize} and {\it approximate} MTTKRP calculations. 
The parallelization strategies have been developed both for dense tensors on GPUs~\cite{hayashi2017shared} and distributed memory systems~\cite{ballard2018parallel,liavas2017nesterov}, and for sparse tensors on GPUs~\cite{nisa2019load} and distributed memory systems~\cite{li2017model,smith2015splatt,smith2016medium,kaya2018parallel}.
The communication lower bounds for a single dense MTTKRP computation has been discussed in~\cite{ballard2020general,ballard2018communication}. 
Approximate CP-ALS algorithms typically accelerate MTTKRP by replacing it with a cheaper approximate computation.
For example, \cite{battaglino2017practical} calculates MTTKRP based on random sampling, and \cite{ma2018accelerating} approximates MTTKRP based on pairwise perturbation. These methods are shown to be efficient on various datasets.

Pairwise perturbation (PP)~\cite{ma2018accelerating} uses perturbative  corrections to the subproblems rather than recomputing the tensor contractions. It has two steps, the \textit{initialization step}, where the PP operators are calculated and amortized, and the \textit{approximated step}, where the PP operators are leveraged to approximate the MTTKRP. For an order $N$ tensor with dimension size $s$ and CP rank $R$, the approximated step reduces the per-sweep computational cost from $\bigo(s^NR)$ to $\bigo(N^2(s^2R+R^2))$. PP has been shown to be accurate when the factor matrices are changing little across sweeps, occurring when CP-ALS approaches convergence. Therefore, PP can speed-up the process to achieve high fitness (the similarity between the input and the reconstructed tensor), making PP useful for applications in scientific computing, such as quantum chemistry.

Despite the advantages of PP for many synthetic and real tensors shown in~\cite{ma2018accelerating}, 
effective parallelization of PP is challenging.
Previous implementations of the algorithm~\cite{ma2018accelerating} used
the parallelism in the tensor contraction kernels in the Cyclops~\cite{solomonik2014massively} library as well as the linear system solve routines in ScaLAPACK~\cite{Dongarra:1997:SUG:265932}. For a specific tensor contraction, Cyclops redistributes input tensors to mappings that are efficient for the contraction, while bringing potential communication overheads between two consecutive contractions in MTTKRP.
We explore parallelization strategies that reduce communication by reorganizing the overall PP algorithm as opposed to just parallelizing each kernel therein.

In this paper, we propose two new parallel algorithms to accelerate \mbox{MTTKRP} calculations in CP-ALS for \textit{dense} tensors. 
These two algorithms are both efficient in communication, and are computationally more efficient than the state-of-art dimension tree based CP-ALS algorithm.

First, we propose the \textit{multi-sweep dimension tree} (MSDT) algorithm, which requires the TTM between an order-$N$ input tensor with dimension size $s$ and the first-contracted input matrix once every $\frac{N-1}{N}$ sweeps and reduce the leading per-sweep computational cost of a rank-$R$ CP-ALS to $2\frac{N}{N-1}s^NR$. This algorithm can produce exactly the same results as the standard dimension tree, 
i.e., it has no accuracy loss.
Leveraging a parallelization strategy similar to
previous work~\cite{ballard2018parallel,eswar2019planc} that performs the dimension tree calculations locally, our benchmark results show a speed-up of $1.25$X compared to the state-of-art dimension tree running on 1024 processors. 

Second, we propose a communication-efficient pairwise perturbation algorithm. The implementation also uses a parallelization strategy similar to \cite{ballard2018parallel,eswar2019planc}, and 
reduces the communication cost by
constructing the first-order local PP operators in the PP initialization step as well as constructing the local MTTKRP approximations in the PP approximated step. Our benchmark results show that the PP approximated step achieves a speed-up of $1.94$X compared to the state-of-art dimension tree running on 1024 processors. 

In summary, this paper makes the following contributions:
\begin{itemize}
    \item we propose a new multi-sweep dimension tree algorithm that reduces the leading computational cost of a rank-$R$ CP-ALS to $2\frac{N}{N-1}s^NR$ for each ALS sweep,
    \item we propose a communication-efficient pairwise perturbation algorithm,
    \item our implementations obtain efficient parallel scaling, and achieve at least 1.52X speed-ups compared to the state-of-art CP-ALS parallel algorithms on real datasets.
\end{itemize}

\section{Background}
\label{sec:background}
\subsection{Notations and Definitions}
We use both element-wise and specialized tensor algebra 
notation~\cite{kolda2009tensor}.
Vectors are denoted with bold lowercase Roman letters (e.g., $\vcr{v}$), matrices are denoted with bold uppercase Roman letters (e.g., $\mat{M}$), and tensors are denoted with bold calligraphic fonts (e.g., $\tsr{T}$). 
An order $N$ tensor corresponds to an $N$-dimensional array. 
Elements of vectors, matrices, and tensors are denotes in parentheses, e.g., $\vcr{v}(i)$ for a vector $\vcr{v}$, $\mat{M}(i,j)$ for a matrix $\mat{M}$, and $\tsr{T}(i,j,k,l)$ for an order 4 tensor $\tsr{T}$.
The $i$th column of $\mat{M}$ is denoted by $\mat{M}(:,i)$. Parenthesized superscripts are used to label different vectors, matrices and tensors (e.g. $\tsr{T}^{(1)}$ and $\tsr{T}^{(2)}$ are unrelated tensors).

The pseudo-inverse of matrix $\mat{A}$ is denoted with $\mat{A}^\dagger$. The Hadamard product of two matrices is denoted with $*$. The outer product of two or more vectors is denoted with $\circ$. The Kronecker product of two vectors/matrices is denoted with $\otimes$.
For matrices $\mat{A}\in \mathbb{R}^{I\times K}$ and $\mat{B}\in \mathbb{R}^{J\times K}$, their Khatri-Rao product resulting in a matrix of size $(IJ)\times K$ defined by
$
   \mat{A}\odot \mat{B} = [\vcr{a}_1\otimes \vcr{b}_1,\ldots, \vcr{a}_K\otimes \vcr{b}_K] .
$
The mode-$n$ TTM of an order $N$ tensor $\tsr{T} \in \mathbb{R}^{s_1\times \cdots \times s_N}$ with a matrix $ \textbf{A}\in \mathbb{R}^{J\times s_n}$ is denoted by $\tsr{T}\times_n \mat{A}$, whose output size is $s_1\times\cdots\times s_{n-1}\times J\times s_{n+1}\times\cdots\times s_N$.
Matricization is the process of unfolding a tensor into a matrix. The mode-$n$ matricized version of $\tsr{T}$ is denoted by $\textbf{T}_{(n)}\in \mathbb{R}^{s_n\times K}$ where $K=\prod_{m=1,m\neq n}^N s_m$. 
We generalize this notation to define the unfoldings of a tensor $\tsr{T}$ with dimensions $s_1 \times \cdots \times s_N$ into an order $M+1$ tensor, $\tsr{T}_{(i_1,\ldots,i_M)}\in \mathbb{R}^{s_{i_1} \times \cdots \times s_{i_M}\times K}$, where $K=\prod_{j\in \{1,\ldots, N\} \setminus \{i_1,\ldots, i_M\}} s_j$. For instance, if $\tsr{T}$ is an order 4 tensor,
$\tsr{T}(j,k,l,m) = \tsr{T}_{(1,3)}(j,l,k+(m-1)s_2).$

We used calligraphic fonts (e.g., $\proc{P}$) to denote the tensor representing a logical multidimensional processor grid. 
Similar to the representation of tensor elements, the index of a specific processor in the grid is denoted in parentheses, e.g. $\proc{P}(i,j)$ denotes one processor indexed by $i,j$ in the 2-dimensional grid. 
For a processor grid $\proc{P}$ with size $I_1\times\cdots\times I_N$ and a tensor $\tsr{T}\in\R^{s_1\times\cdots\times s_N}$, 
let $x=(x_1,\ldots,x_N)$ denote the index on $\proc{P}$, we use $\tsr{T}_{\proc{P}(x)}$ to denote the local tensor residing on a processor indexed by $x$. 
The size of the local tensor will be ${\lceil\frac{s_1}{I_1}\rceil\times\cdots\times \lceil\frac{s_N}{I_N}\rceil}$. When $\frac{s_i}{I_i}$ is not an integer, paddings will be added to the tensor.

\subsection{CP Decomposition with ALS}
\label{sec:bg-cp}

\begin{algorithm}[t]
    \caption{\textbf{CP-ALS}: ALS for CP decomposition}
\label{alg:cp_als}
\begin{algorithmic}[1]
\small
\STATE{
\textbf{Input: }Tensor $\tsr{T}\in\mathbb{R}^{s_1\times\cdots s_N}$, 
stopping criteria {$\Delta$}, rank $R$
}
\STATE{Initialize $[\![ \mat{A}^{(1)}, \ldots , \mat{A}^{(N)} ]\!]$ with $\mat{A}^{(i)}\in\R^{s_i\times R}$ as uniformly distributed random matrices within $[0,1]$, $ \mat{S}^{(i)} \leftarrow\mat{A}^{(i)T}\mat{A}^{(i)}$ for $i\in\{1,\ldots,N\}$
}

\STATE{$r\leftarrow 1, r_{\text{old}} \leftarrow 0$ \Comment{Initialize the relative residual}}
\WHILE{$|r - r_{\text{old}}|>{\Delta}$}
\FOR{\texttt{$i\in \inti{1}{N} $}} \label{line:cpals_for}
\STATE\label{line3}{$\boldsymbol{\Gamma}^{(i)}\leftarrow\textbf{S}^{(1)}\ast\cdots\ast \textbf{S}^{(i-1)}\ast \textbf{S}^{(i+1)}\ast\cdots\ast \textbf{S}^{(N)} $}
\STATE{Update $ \textbf{\M}^{(i)}$ via dimension tree in Section~\ref{sec:bg-dt}}
\STATE{$   \textbf{A}^{(i)} \leftarrow \textbf{\M}^{(i)}\boldsymbol{\Gamma}^{(i)}{}^{\dagger} $}
\vspace{.01in}
\STATE{$   \textbf{S}^{(i)} \leftarrow {\textbf{A}^{(i)}{}^T}\textbf{A}^{(i)} $}
\ENDFOR \label{line:cpals_endfor}

\STATE{$r_{\text{old}} = r$}
\STATE{Update $r$ based on Equation~\eqref{eq:residual}}

\ENDWHILE
\RETURN $[\![ \textbf{A}^{(1)}, \ldots , \textbf{A}^{(N)} ]\!]$
\end{algorithmic}
\end{algorithm}

The goal of the CP tensor decomposition is to minimize the following objective function:
   \[
f(\mat{A}^{(1)}, \ldots , \mat{A}^{(N)}) \defeq
 \frac{1}{2}\|\tsr{T}-[\![ \mat{A}^{(1)}, \cdots , \mat{A}^{(N)} ]\!]\|_F^2, 
   \]
where 
\[
    [\![ \mat{A}^{(1)}, \cdots , \mat{A}^{(N)} ]\!]
    :=\sum_{r=1}^{R} \mat{A}^{(1)}(:,r)\circ \cdots \circ \mat{A}^{(N)}(:,r).
    \]
CP-ALS alternates among subproblems for each of the factor matrices $\mat{A}^{(n)}$. Each subproblem is quadratic, with the optimality condition being setting the subproblem gradient
$\frac{\partial f}{\partial \mat{A}^{(n)}}$
to zero, resulting in the update expression,
     \[
     \mat{A}^{(n)}_{\text{new}}\boldsymbol{\Gamma}^{(n)}= \mat{T}_{(n)}\mat{P}^{(n)},
     \]  
where the matrix $\mat{P}^{(n)}\in \mathbb{R}^{I_n \times R}$, with $I_n =
\prod_{i=1,i\neq n}^{N}s_i
$, is formed by Khatri-Rao products of the other factor matrices,
\[
    \mat{P}^{(n)}=\mat{A}^{(N)} \odot \cdots \odot  \mat{A}^{(n+1)}  \odot  \mat{A}^{(n-1)} \odot \cdots \odot \mat{A}^{(1)},
\]
and $\mat{\Gamma}\in\mathbb{R}^{R\times R}$ is computed via a chain of Hadamard products,
\begin{equation}
\boldsymbol{\Gamma}^{(n)}=\textbf{S}^{(1)}\ast\cdots\ast \textbf{S}^{(n-1)} \ast \textbf{S}^{(n+1)}\ast\cdots\ast \textbf{S}^{(N)},
\label{eq:gram}
\end{equation}
\[\text{with each} \quad \textbf{S}^{(i)} = \textbf{A}^{(i)T}\textbf{A}^{(i)}.\]
The MTTKRP computation $\mat{M}^{(n)}=\mat{T}_{(n)}\mat{P}^{(n)}$ is the main computational bottleneck of CP-ALS.
 The computational cost of MTTKRP is $\bigo(s^NR)$ if $s_n=s$ for all $n\in\inti{1}{N}$. 
 The naive implementation of CP-ALS for a dense tensor calculates $N$ MTTKRPs for each ALS sweep, leading to the cost of $\bigo(Ns^NR)$.
With the dimension tree algorithm, the computational complexity for all the MTTKRP calculations in one ALS iteration is $4s^{N}R$ to leading order in $s$. The dimension tree algorithm will be detailed in Section~\ref{sec:bg-dt}.
Algorithm~\ref{alg:cp_als} presents the CP-ALS procedure described above, performing iterations until the relative decomposition residual of the neighboring sweeps is sufficiently small.
Let $\Tilde{\tsr{T}}$ denote the tensor reconstructed by the factor matrices, the relative residual norm is defined as
\begin{equation}
    r = \frac{\|\tsr{T} - \Tilde{\tsr{T}}\|_F}{\|\tsr{T}\|_F}.
    \label{eq:residual-def}
\end{equation}
As is shown~\cite{ballard2018parallel,liavas2017nesterov,smith2016medium}, $r$ can be calculated efficiently via
\begin{equation}
\small{
    \label{eq:residual}
    r =
\frac{
\sqrt{
\|\tsr{T}\|_F + 
\langle \boldsymbol{\Gamma}^{(N)} , \mat{A}^{(N)T}\mat{A}^{(N)}  \rangle
-2\langle \mat{M}^{(N)} , \mat{A}^{(N)}  \rangle
}}{\|\tsr{T}\|_F}, 
}\end{equation}
assuming that terms $\|\tsr{T}\|_F, \boldsymbol{\Gamma}^{(N)}, \mat{M}^{(N)}$ are all amortized before the residual calculations.

\subsection{The Dimension Tree Algorithm}
\label{sec:bg-dt}

\newsavebox{\dtfigbox}
\begin{lrbox}{\dtfigbox}
\begin{tikzpicture}[>=latex',scale=0.41]
    \tikzstyle{n} = [draw,shape=ellipse,minimum size=1.5em,
                        inner sep=0pt,fill=violet!20]
\node (X) at (137.0bp,234.0bp) [draw,ellipse,n] {$\tsr{T}$};
  \node (M123) at (96.0bp,162.0bp) [draw,ellipse,n] {\tiny{$\tsr{M}^{(1,2,3)}$}};
  \node (M234) at (179.0bp,162.0bp) [draw,ellipse,n] {\tiny{$\tsr{M}^{(2,3,4)}$}};
  \node (M12) at (96.0bp,90.0bp) [draw,ellipse,n] {\tiny{$\tsr{M}^{(1,2)}$}};
  \node (M34) at (179.0bp,90.0bp) [draw,ellipse,n] {\tiny{$\tsr{M}^{(3,4)}$}};
  \node (M1) at (27.0bp,18.0bp) [draw,ellipse,n] {\small{$\mat{M}^{(1)}$}};
  \node (M2) at (99.0bp,18.0bp) [draw,ellipse,n] {\small{$\mat{M}^{(2)}$}};
  \node (M3) at (176.0bp,18.0bp) [draw,ellipse,n] {\small{$\mat{M}^{(3)}$}};
  \node (M4) at (248.0bp,18.0bp) [draw,ellipse,n] {\small{$\mat{M}^{(4)}$}};
  \draw [->] (X) ..controls (122.54bp,208.32bp) and (116.38bp,197.8bp)  .. (M123);
  \draw [->] (X) ..controls (151.67bp,208.56bp) and (158.12bp,197.8bp)  .. (M234);
  \draw [->] (M123) ..controls (96.0bp,135.98bp) and (96.0bp,126.71bp)  .. (M12);
  \draw [->] (M234) ..controls (179.0bp,135.98bp) and (179.0bp,126.71bp)  .. (M34);
  \draw [->] (M12) ..controls (72.081bp,64.735bp) and (59.223bp,51.69bp)  .. (M1);
  \draw [->] (M12) ..controls (97.072bp,63.983bp) and (97.469bp,54.712bp)  .. (M2);
  \draw [->] (M34) ..controls (177.93bp,63.983bp) and (177.53bp,54.712bp)  .. (M3);
  \draw [->] (M34) ..controls (202.92bp,64.735bp) and (215.78bp,51.69bp)  .. (M4);
    \node[left=3em] at (M1)  (l3) {Level 3};
    \node at (l3 |- M12) (l2){Level 2};
    \node at (l3 |- M123) (l1) {Level 1};
\end{tikzpicture}
\end{lrbox}

\newsavebox{\ppfigbox}
\begin{lrbox}{\ppfigbox}
\pgfdeclarelayer{background}
\pgfdeclarelayer{foreground}
\pgfsetlayers{background,main,foreground}
\begin{tikzpicture}[>=latex',scale=0.41]
    \tikzstyle{n} = [draw,shape=ellipse,minimum size=1.5em,
                        inner sep=0pt,fill=violet!20]
\node (X) at (265.95bp,234.0bp) [draw,ellipse,n] {$\tsr{T}$};
  \node (M123) at (150.95bp,162.0bp) [draw,ellipse,n] {\tiny{$\tsr{M}_p^{(1,2,3)}$}};
  \node (M134) at (265.95bp,162.0bp) [draw,ellipse,n] {\tiny{$\tsr{M}_p^{(1,3,4)}$}};
  \node (M234) at (380.95bp,162.0bp) [draw,ellipse,n] {\tiny{$\tsr{M}_p^{(2,3,4)}$}};
  \node (M12) at (101.95bp,90.0bp) [draw,ellipse,n] {\tiny{$\tsr{M}_p^{(1,2)}$}};
  \node (M13) at (27.948bp,90.0bp) [draw,ellipse,n] {\tiny{$\tsr{M}_p^{(1,3)}$}};
  \node (M23) at (175.95bp,90.0bp) [draw,ellipse,n] {\tiny{$\tsr{M}_p^{(2,3)}$}};
  \node (M14) at (249.95bp,90.0bp) [draw,ellipse,n] {\tiny{$\tsr{M}_p^{(1,4)}$}};
  \node (M34) at (323.95bp,90.0bp) [draw,ellipse,n] {\tiny{$\tsr{M}_p^{(3,4)}$}};
  \node (M24) at (397.95bp,90.0bp) [draw,ellipse,n] {\tiny{$\tsr{M}_p^{(2,4)}$}};
  \node (M1) at (101.95bp,18.0bp) [draw,ellipse,n] {\small{$\mat{M}^{(1)}$}};
  \node (M2) at (246.95bp,18.0bp) [draw,ellipse,n] {\small{$\mat{M}^{(2)}$}};
  \node (M3) at (174.95bp,18.0bp) [draw,ellipse,n] {\small{$\mat{M}^{(3)}$}};
  \node (M4) at (323.95bp,18.0bp) [draw,ellipse,n] {\small{$\mat{M}^{(4)}$}};
  \draw [->] (X) ..controls (228.76bp,210.37bp) and (201.74bp,193.92bp)  .. (M123);
  \draw [->] (X) ..controls (265.95bp,207.98bp) and (265.95bp,198.71bp)  .. (M134);
  \draw [->] (X) ..controls (303.13bp,210.37bp) and (330.15bp,193.92bp)  .. (M234);
  \draw [->] (M123) ..controls (133.37bp,135.89bp) and (125.56bp,124.74bp)  .. (M12);
  \draw [->] (M123) ..controls (108.88bp,137.06bp) and (79.072bp,120.1bp)  .. (M13);
  \draw [->] (M123) ..controls (159.86bp,136.06bp) and (163.33bp,126.33bp)  .. (M23);
  \draw [->] (M134) ..controls (260.29bp,136.26bp) and (258.14bp,126.82bp)  .. (M14);
  \draw [->] (M134) ..controls (286.4bp,136.31bp) and (296.15bp,124.55bp)  .. (M34);
  \draw [->] (M234) ..controls (386.96bp,136.26bp) and (389.25bp,126.82bp)  .. (M24);
  \draw [->,dashed,blue!60] (M12) ..controls (101.95bp,63.983bp) and (101.95bp,54.712bp)  .. (M1);
  \draw [->,dashed,blue!60] (M12) ..controls (148.25bp,66.647bp) and (188.51bp,47.209bp)  .. (M2);
  \draw [->,dashed,blue!60] (M13) ..controls (53.768bp,64.575bp) and (67.842bp,51.262bp)  .. (M1);
  \draw [->,dashed,blue!60] (M13) ..controls (74.888bp,66.647bp) and (115.71bp,47.209bp)  .. (M3);
  \draw [->,dashed,blue!60] (M14) ..controls (202.83bp,66.712bp) and (161.01bp,46.935bp)  .. (M1);
  \draw [->,dashed,blue!60] (M14) ..controls (275.77bp,64.575bp) and (289.84bp,51.262bp)  .. (M4);
  \draw [->,dashed,blue!60] (M23) ..controls (200.56bp,64.735bp) and (213.79bp,51.69bp)  .. (M2);
  \draw [->,dashed,blue!60] (M23) ..controls (175.59bp,63.983bp) and (175.46bp,54.712bp)  .. (M3);
  \draw [->,dashed,blue!60] (M24) ..controls (350.12bp,66.83bp) and (306.98bp,46.829bp)  .. (M2);
  \draw [->,dashed,blue!60] (M24) ..controls (372.13bp,64.575bp) and (358.05bp,51.262bp)  .. (M4);
  \draw [->,dashed,blue!60] (M34) ..controls (279.07bp,67.988bp) and (242.51bp,50.856bp)  .. (210.95bp,36.0bp) .. controls (209.17bp,35.161bp) and (207.33bp,34.298bp)  .. (M3);
  \draw [->,dashed,blue!60] (M34) ..controls (323.95bp,63.983bp) and (323.95bp,54.712bp)  .. (M4);
    \node[left=1em] at (M1)  (l3) {};
    \node at (l3 |- M12) (l2){};
    \node at (l3 |- M123) (l1) {};

    \begin{pgfonlayer}{background}
        \draw[rounded corners=2em,line width=3em,blue!20,cap=round]
                (M13.west) -- (M24.east);
    \end{pgfonlayer}
\end{tikzpicture}
\end{lrbox}

\begin{figure*}[]
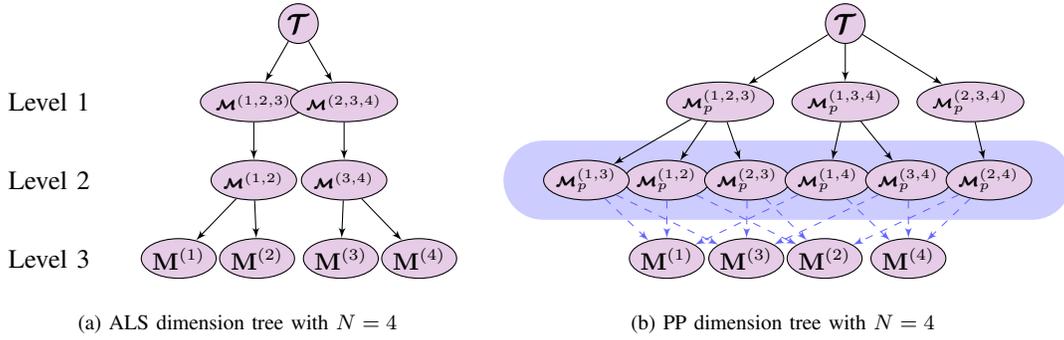

\centering
\captionsetup[subfloat]{farskip=2pt,captionskip=13pt}
\subfloat[ALS dimension tree with $N=4$]{\usebox{\dtfigbox}
\label{subfig:dt}
}
\subfloat[PP dimension tree with $N=4$]{\usebox{\ppfigbox}
\label{subfig:pp}
}

\caption{Dimension trees for ALS and pairwise perturbation. The blue region in (b) denotes the PP operators.}
\end{figure*}

In each CP-ALS sweep, the TTM and TTV operations for MTTKRP calculations can be amortized and reused.  
Such amortization strategies are referred to as dimension tree algorithms. 
A dimension tree data structure partitions the mode indices of  an order $N$ tensor hierarchically and constructs the intermediate tensors accordingly~\cite{kaya2019computing,phan2013fast,ballard2018parallel}. The root of the tree corresponds to the input tensor and the leaves consist of all the $N$ factor matrices.
It is assumed in the literature that each ALS sweep uses the same tree.

It has been shown in \cite{kaya2019computing} that for CP decomposition, an optimal dimension tree must be binary. Therefore, our analysis will focus on the binary trees. 
We illustrate one dimension tree for $N=4$ in Figure~\ref{subfig:dt}. 
The first level contractions (contractions between the input tensor and one factor matrix) are done via TTM.
For an equidimensional tensor with size $s$ and rank $R$,
these contractions have a cost of $\bigo(s^NR)$ and are generally the most time-consuming part of ALS. Other contractions (transforming one intermediate into another intermediate) are done via batched TTV (also called multi-TTV/mTTV), and the complexity of an $i$th level contraction, where $i<N$,  is $\bigo(s^{N+1-i}R)$. Because two first level contractions are necessary for the construction of the dimension tree, as is illustrated in Figure~\ref{subfig:dt}, to calculate all the MTTKRP results in one ALS iteration, to leading order in $s$, the computational complexity is $4s^{N}R.$

\subsection{The Pairwise Perturbation Algorithm}
\label{sec:bg-pp}

\begin{algorithm}[]
    \caption{\textbf{PP-CP-ALS}: Pairwise perturbation for CP-ALS}
\label{alg:cp_als_pp}
\begin{algorithmic}[1]
\small
\STATE{\textbf{Input:} tensor $\tsr{T}\in\mathbb{R}^{s_1\times\cdots\times s_N}$, 
       stopping criteria $\Delta$, PP tolerance $\epsilon<1$}
\STATE{Initialize $[\![ \textbf{A}^{(1)}, \ldots , \textbf{A}^{(N)} ]\!]$ with $\mat{A}^{(i)}\in\R^{s_i\times R}$ as uniformly distributed random matrices within $[0,1]$, $d\mat{A}^{(i)}\leftarrow \mat{A}^{(i)}$, $ \mat{S}^{(i)} \leftarrow\mat{A}^{(i)T}\mat{A}^{(i)}$ for $i\in\{1,\ldots,N\}$
}
\STATE{$r\leftarrow 1, r_{\text{old}} \leftarrow 0$ \Comment{Initialize the relative residual}}
\WHILE{$|r - r_{\text{old}}|>{\Delta}$}
  \IF{$\forall ~i \in \inti{1}{N}, {\fnrm{d\mat{A}^{(i)}}}<\epsilon{\fnrm{\mat{A}^{(i)}}}$}
    \FOR{\texttt{$i\in \inti{1}{N} $}}
      \STATE{$\mat{A}_p^{(i)} \gets \mat{A}^{(i)}$, $d\mat{A}^{(i)} \gets \mat{O}$}
    \ENDFOR
    \STATE{Compute $\tsr{M}^{(i,n)}_p,\mat{M}^{(n)}_p$ for $i,n\in \inti{1}{N}$ via dimension tree \label{line:pp_init_step}}
    \WHILE{$\forall ~i \in \inti{1}{N}, \fnrm{d\mat{A}^{(i)}}<\epsilon\fnrm{\mat{A}^{(i)}}$}
      \FOR{\texttt{$j\in \inti{1}{N} $}}
        \STATE{$\boldsymbol{\Gamma}^{(j)}\leftarrow\mat{S}^{(1)}\ast\cdots\ast \mat{S}^{(j-1)}\ast \mat{S}^{(j+1)}\ast\cdots\ast \mat{S}^{(N)} $}
        \STATE{
        Update $\Tilde{\mat{M}}^{(j)}$ based on Equation~\eqref{eq:ppupdate}
         }
        \STATE{$   \mat{A}^{(j)} \leftarrow \Tilde{\mat{M}}^{(j)}\boldsymbol{\Gamma}^{(j)}{}^{\dagger} $}
        \STATE{$d\mat{A}^{(j)} = \mat{A}^{(j)} -\mat{A}^{(j)}_p$, $   \mat{S}^{(j)} \leftarrow {\mat{A}^{(j)}}{}^T\mat{A}^{(j)} $}
      \ENDFOR
    \ENDWHILE
  \ENDIF \label{line:pp_endif}
  \STATE{Perform a regular ALS sweep as in Algorithm~\ref{alg:cp_als} (line \ref{line:cpals_for}-\ref{line:cpals_endfor}), taking $d\mat{A}^{(i)}$ as the update of $\mat{A}^{(i)}$ in one sweep for each $i\in\inti{1}{N}$ \label{line:regular_als}}
  \STATE{$r_{\text{old}} = r$}
  \STATE{Update $r$ based on Equation~\eqref{eq:residual}}
\ENDWHILE
\RETURN $[\![ \mat{A}^{(1)}, \ldots , \mat{A}^{(N)} ]\!]$
\end{algorithmic}
\end{algorithm}

Before the introduction of PP, we define the partially contracted MTTKRP intermediates $\tsr{M}^{(i_1,i_2,\ldots,i_m)}$ as follows,
\begin{equation}
\small{
\tsr{M}^{(i_1,i_2,\ldots,i_m)} = \tsr{T}_{(i_1,i_2,\ldots,i_m)}\bigodot_{j\in {\inti{1}{N}}\setminus\{i_1,i_2,\ldots, i_m\}}\mat{A}^{(j)},}
\label{eq:tensors-cp} 
\end{equation}
where $\tsr{M}^{(1,\ldots, N)}$ is the input tensor $\tsr{T}$. Let $\mat{A}_p^{(n)}$ denote the $\mat{A}^{(n)}$ calculated with regular ALS at some number of sweeps prior to the current one,
we also define $\tsr{M}_p^{(i_1,i_2,\ldots,i_m)}$ 
in the same way as $\tsr{M}^{(i_1,i_2,\ldots,i_m)}$ in Equation~\eqref{eq:tensors-cp}, except that $\tsr{T}$ is contracted with $\mat{A}_p^{(j)}$ for $j\in \inti{1}{N} \setminus \{i_1,i_2,\ldots,i_m\}$.

Pairwise perturbation (PP) uses perturbative  corrections to the subproblems rather than recomputing the tensor contractions and contains two steps. 
The initialization step calculates the PP operators $\tsr{M}_p^{(i,j)}$ for $ \forall i,j\in\{1,\ldots, N\}, i<j$. 
Similar to CP-ALS, computation of these operators can also benefit from dimension trees.
Figure~\ref{subfig:pp} describes the PP dimension tree for $N=4$.
In the PP dimension tree, ${l+1 \choose 2}$ tensors $\tsr{M}_p^{(i,j,j+1,\ldots, j+N-l-1)}, \forall i,j\in \inti{1}{l+1}, i<j$ are calculated at the $l$th level.
The construction of PP operators with the dimension tree costs $4s^NR$ to the leading order, which is computationally the same expensive as the ALS dimension tree algorithm\footnote{Note that despite 3 first level intermediates are needed, only two of them need to be recalculated, and the remaining one (e.g. $\tsr{M}_p^{(2,3,4)}$ in Figure~\ref{subfig:pp}) can be amortized from the last DT sweep.}.

The approximated step uses the PP operators to approximate the MTTKRP $\mat{M}^{(n)}$ with $\Tilde{\mat{M}}^{(n)}$ as follows,
\begin{equation}
    \Tilde{\mat{M}}^{(n)}
    =\mat{M}_p^{(n)} + \sum_{i=1,i\neq n}^{N}\mat{U}^{(n,i)}
    + \mat{V}^{(n)},
   \label{eq:ppupdate}
\end{equation}
\begin{equation}
\text{where} \quad    \mat{U}^{(n,i)}(x,k) = \sum_{y=1}^{s_i}\tsr{M}_p^{(n,i)}(x,y,k) d\mat{A}^{(i)}(y,k),
\label{eq:U_ni}
\end{equation}
\vspace{-3mm}
\begin{equation}
\small{
    \mat{V}^{(n)} = \mat{A}^{(n)} \bigg( 
    \sum_{i,j=1\neq n, i<j}^{N}
    d\mat{S}^{(i)} * d\mat{S}^{(j)} * \bigast_{k=1, k\neq i,j,n}^{N}\mat{S}^{(k)} \bigg),
}
\label{eq:V_nij}
\end{equation}
\begin{equation}
\text{and} \quad  d\mat{S}^{(i)} = \mat{A}^{(i)T} d\mat{A}^{(i)}.
\label{eq:ds}
\end{equation}
Terms $\mat{U}^{(n,i)}$ are the first-order corrections computed via
the PP operators, and the term $\mat{V}^{(n)}$ is the second-order correction to lower the error to a greater extent.
Given $\tsr{M}_p^{(n,i)}$ and $\mat{M}_p^{(n)}$, calculation of $\Tilde{\mat{M}}^{(n)}$ for $n\in \inti{1}{N}$ requires $2N(Ns^2R+NR^2 + sR^2)$ operations overall. Algorithm~\ref{alg:cp_als_pp} presents the PP-CP-ALS method described above. We direct readers to Section 3.1 in reference~\cite{ma2018accelerating} for an illustration of the algorithm on order 3 tensors.

\subsection{Parallel CP-ALS}
\label{subsec:par-als}
\begin{algorithm}
\caption{\textbf{Par-CP-ALS}: Parallel CP-ALS}
\label{alg:Par-CP-ALS}
\begin{algorithmic}[1]
\small
\STATE{
\textbf{Input: }Processor grid $\proc{P}$ with dimension $
I_1\times\cdots\times I_N$, where $I=\prod_{i=1}^N I_i$, tensor $\tsr{T}\in\mathbb{R}^{s_1\times\cdots s_N}$ distributed over $\proc{P}$
}
\STATE{$\proc{Q}\leftarrow \proc{P}$ reshaped to a 2-d array with dimension $I\times 1$}
\STATE{\Comment{$x$ denotes one processor in the grid whose index is $(x_1,\ldots,x_N)$ in $\proc{P}$, and the index is $(x, 1)$ in $\proc{Q}$}}
\FOR{\texttt{$i\in \inti{2}{N} $}}
	\STATE Initialize $\mb{A}{i}_{\proc{Q}(x)}\in \R^{\lceil \frac{s_i}{I}\rceil\times R}$ 
	\STATE $\mat{S}^{(i)}_{\proc{Q}(x)} \leftarrow \mat{A}^{(i)T}_{\proc{Q}(x)}\mt{A}{i}{\proc{Q}(x)}$
	\STATE $\mat{S}^{(i)}_{\proc{Q}(x)} \leftarrow \text{All-Reduce}(\mat{S}_{\proc{Q}(x)},\textsc{All-Procs})$ \label{line:all-reduce1}
	\STATE $\mb{A}{i}_{\proc{P}(x)} \leftarrow \text{All-Gather}(\mb{A}{i}_{\proc{Q}(x)},\textsc{Proc-Slice}(\proc{P}_{(i)}(x_i, :))$\label{line:all-gather1}
\ENDFOR
\WHILE{Stopping criteria not reached}
    \FOR{\texttt{$i\in \inti{1}{N} $}}
        \STATE{$\boldsymbol{\Gamma}^{(i)}_{\proc{Q}(x)}\leftarrow\mat{S}^{(1)}_{\proc{Q}(x)}\ast\cdots\ast \mat{S}^{(i-1)}_{\proc{Q}(x)}\ast \mat{S}^{(i+1)}_{\proc{Q}(x)}\ast\cdots\ast \mat{S}^{(N)}_{\proc{Q}(x)} $}
        \STATE{$ \mat{\M}^{(i)}_{\proc{P}(x)}\leftarrow \text{Local-MTTKRP}(\tsr{T}_{\proc{P}(x)}, \{\mt{A}{1}{\proc{P}(x)},\ldots,\mt{A}{N}{\proc{P}(x)}\}, i)$ via dimension tree in Section~\ref{sec:bg-dt}
        \label{line:local-mttkrp}
        }
        \STATE{$ \mat{\M}^{(i)}_{\proc{Q}(x)}\leftarrow 
        \text{Reduce-Scatter}(\mat{\M}^{(i)}_{\proc{P}(x)}, \textsc{Proc-Slice}(\proc{P}_{(i)}(x_i, :)))
        $
        \label{line:reduce-scatter}
        }
        \STATE{$  \mt{A}{1}{\proc{Q}(x)} \leftarrow \mat{M}^{(i)}_{\proc{Q}(x)}\boldsymbol{\Gamma}_{\proc{Q}(x)}^{(i)\dagger} $}
        \vspace{.01in}
        \STATE{$   \mat{S}^{(i)}_{\proc{Q}(x)} \leftarrow \mat{A}_{\proc{Q}(x)}^{(i)T}\mat{A}^{(i)}_{\proc{Q}(x)} $}
    	\STATE $\mat{S}^{(i)}_{\proc{Q}(x)} \leftarrow \text{All-Reduce}(\mat{S}_{\proc{Q}(x)},\textsc{All-Procs})$ \label{line:all-reduce2}
    	\STATE $\mb{A}{i}_{\proc{P}(x)} \leftarrow \text{All-Gather}(\mb{A}{i}_{\proc{Q}(x)},\textsc{Proc-Slice}(\proc{P}_{(i)}(x_i, :))$ \label{line:all-gather}
	\ENDFOR 
\ENDWHILE
\RETURN $[\![ \mat{A}^{(1)}, \ldots , \mat{A}^{(N)} ]\!]$.
\end{algorithmic}
\end{algorithm}

We use the BSP $\alpha-\beta-\gamma$ model  for parallel cost analysis~\cite{valiant1990bridging,thakur2005optimization}. In addition to considering the communication cost of sending data among processors (horizontal communication cost), we use another parameter, $\nu$, to measure the cost of transferring data between slow memory and cache (vertical communication cost)~\cite{solomonik2017communication}.
$\alpha$ represents the cost of sending/receiving a single message, $\beta$ represents the cost of moving a single word among processors,
$\nu$ represents the cost of moving a single word between the main memory and cache,
and $\gamma$ represents the cost to perform one arithmetic operation. We assume $\alpha \gg \beta \gg \gamma$, and $\nu \leq \gamma \cdot \sqrt{H}$, where $H$ is the cache size. We summarize the collective routines on a fully-connected network used in the parallel algorithms, including All-Gather, Reduce-Scatter and All-Reduce, as follows:
\begin{itemize}
    \item $\text{All-Gather}(v, \textsc{Procs})$ collects data $v$ distributed across $P$ processors ($\textsc{Procs}$) and stores the concatenation of all the data with size $n$ redundantly on all processors. Its cost is $\log P \cdot \alpha + n\delta(P) \cdot \beta $, where $\delta(P)=1$ if $P>1$, otherwise $0$.
    \item $\text{Reduce-Scatter}(v, \textsc{Procs})$ sums $v$ distributed across $\textsc{Procs}$ and partitions the result across $\textsc{Procs}$. Its cost is $\log P \cdot \alpha + n\delta(P) \cdot \beta $.
    \item $\text{All-Reduce}(v, \textsc{Procs})$ sums $v$ distributed across $\textsc{Procs}$ and stores the result redundantly on all processors. Its cost is $2\log P \cdot \alpha + 2n\delta(P) \cdot \beta $.
\end{itemize}

Our parallel algorithms for CP-ALS on dense tensors are based on Algorithm~\ref{alg:Par-CP-ALS}, which is introduced in~\cite{ballard2018parallel,eswar2019planc}. The input tensor $\tsr{T}$ with order $N$ is uniformly distributed across an order $N$ processor grid $\proc{P}$, and all the factor matrices are initially distributed such that each processor owns a subset of the rows. The main idea of the algorithm is to 
decompose the MTTKRP into local MTTKRP calculations to reduce the expensive communication cost, e.g.,
if $\mat A^{(1)} = \text{MTTKRP}(\tsr{T}, \mat{A}^{(2)}, \mat{A}^{(3)})$ then
\[
\small
\mat A_{\proc{P}(i,j,k)}^{(1)} = \sum_{j,k} \text{Local-MTTKRP}(\tsr{T}_{\proc{P}(i,j,k)}, \mat{A}^{(2)}_{\proc{P}(i,j,k)}, \mat{A}^{(3)}_{\proc{P}(i,j,k)}),\]
\normalsize
where $i,j,k$ denote the location of a processor in a 3D grid.
Before the MTTKRP calculation, all the factor matrices are redistributed (lines~\ref{line:all-gather1},\ref{line:all-gather}). For the $i$th mode factor matrix $\mat{A}^{(i)}$, all processors having the same $i$th index in the processor grid $\proc{P}$ redundantly own the same $\lceil\frac{s_i}{I_i}\rceil$ rows of $\mat{A}^{(i)}$.
This setup prepares all the local factor matrices $\mt{A}{1}{\proc{P}(x)},\ldots,\mt{A}{N}{\proc{P}(x)}$ necessary for a local MTTKRP on each processor indexed $x$. The local MTTKRP routine (line~\ref{line:local-mttkrp}) independently performs MTTKRP calculations on each processor, and this step requires no communication. The local MTTKRP can be efficiently calculated with the dimension tree techniques. After the local MTTKRP, Reduce-Scatter (line~\ref{line:reduce-scatter}) is performed to sum-up the local contributions and the MTTKRP outputs $\mat{M}^{(i)}$ are distributed such that each processor owns a subset of the rows. In addition, all the Gram matrices $\mat{S}^{(i)}$, once updated, are then distributed redundantly on all the processors via All-Reduce (lines~\ref{line:all-reduce1},\ref{line:all-reduce2}), allowing for the linear systems to be solved locally. 

For each processor, the leading order computational cost is $\frac{4s^NR}{P}$, if the dimension size is $s$ and the state-of-art dimension tree algorithm is used. Three collective routines (lines~\ref{line:reduce-scatter},\ref{line:all-reduce2},\ref{line:all-gather}) are called for each factor matrix update. When $I_1=\cdots=I_N=P^{\frac{1}{N}}$, the horizontal communication cost for each ALS sweep is $\bigo(N\log(P)\cdot\alpha + N(sR/P^{\frac{1}{N}} + R^2)\cdot\beta)$.

Unlike Algorithm~\ref{alg:Par-CP-ALS}, our implementation calculates all the Hadamard products and the linear system solves in a parallel way, leveraging 
a distributed-memory matrix library for solving symmetric positive definite linear systems.
Distributing the work in the solve reduced the computational and bandwidth costs, while raising the latency cost.
Our performance evaluation in Section~\ref{subsec:bench} also includes the PLANC implementation of CP-ALS~\cite{eswar2019planc}, which makes use of a sequential linear system solve.
As we will show, 
the cost of solving linear systems is often small for both approaches,
considering that the MTTKRP calculations are the major bottleneck.

\section{Multi-Sweep Dimension Tree}
\label{sec:pp-msdt}
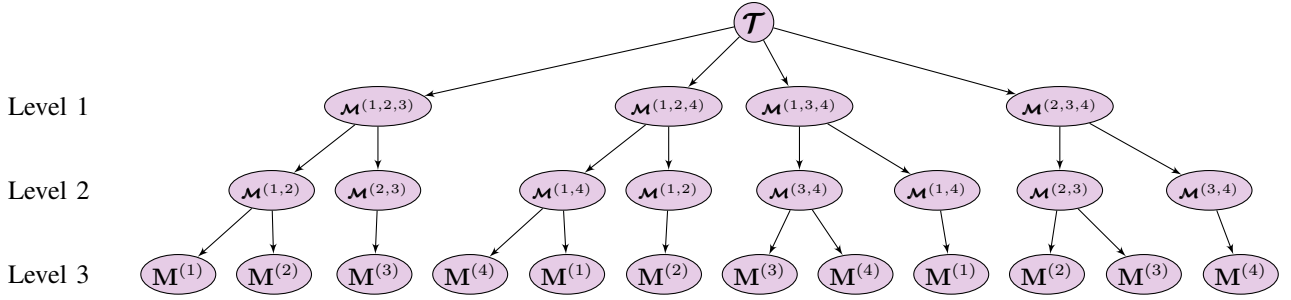
\begin{figure*}[]
\centering
\pgfdeclarelayer{background}
\pgfdeclarelayer{foreground}
\pgfsetlayers{background,main,foreground}

\begin{tikzpicture}[>=latex',scale=0.44]
    \tikzstyle{n} = [draw,shape=ellipse,minimum size=1.5em,
                        inner sep=0pt,fill=violet!20]
\node (X) at (524.85bp,234.0bp) [draw,ellipse,n] {$\tsr{T}$};
  \node (M123) at (202.85bp,162.0bp) [draw,ellipse,n] {\tiny{$\tsr{M}^{(1,2,3)}$}};
  \node (M124) at (451.85bp,162.0bp) [draw,ellipse,n] {\tiny{$\tsr{M}^{(1,2,4)}$}};
  \node (M134) at (563.85bp,162.0bp) [draw,ellipse,n] {\tiny{$\tsr{M}^{(1,3,4)}$}};
  \node (M234) at (786.85bp,162.0bp) [draw,ellipse,n] {\tiny{$\tsr{M}^{(2,3,4)}$}};
  \node (M12_a) at (111.85bp,90.0bp) [draw,ellipse,n] {\tiny{$\tsr{M}^{(1,2)}$}};
  \node (M23_a) at (202.85bp,90.0bp) [draw,ellipse,n] {\tiny{$\tsr{M}^{(2,3)}$}};
  \node (M14_b) at (360.85bp,90.0bp) [draw,ellipse,n] {\tiny{$\tsr{M}^{(1,4)}$}};
  \node (M12_b) at (451.85bp,90.0bp) [draw,ellipse,n] {\tiny{$\tsr{M}^{(1,2)}$}};
  \node (M34_c) at (563.85bp,90.0bp) [draw,ellipse,n] {\tiny{$\tsr{M}^{(3,4)}$}};
  \node (M14_c) at (681.85bp,90.0bp) [draw,ellipse,n] {\tiny{$\tsr{M}^{(1,4)}$}};
  \node (M23_d) at (786.85bp,90.0bp) [draw,ellipse,n] {\tiny{$\tsr{M}^{(2,3)}$}};
  \node (M34_d) at (914.85bp,90.0bp) [draw,ellipse,n] {\tiny{$\tsr{M}^{(3,4)}$}};
  \node (M1_a) at (31.847bp,18.0bp) [draw,ellipse,n] {\small{$\mat{M}^{(1)}$}};
  \node (M2_a) at (113.85bp,18.0bp) [draw,ellipse,n] {\small{$\mat{M}^{(2)}$}};
  \node (M3_a) at (199.85bp,18.0bp) [draw,ellipse,n] {\small{$\mat{M}^{(3)}$}};
  \node (M4_b) at (281.85bp,18.0bp) [draw,ellipse,n] {\small{$\mat{M}^{(4)}$}};
  \node (M1_b) at (364.85bp,18.0bp) [draw,ellipse,n] {\small{$\mat{M}^{(1)}$}};
  \node (M2_b) at (447.85bp,18.0bp) [draw,ellipse,n] {\small{$\mat{M}^{(2)}$}};
  \node (M3_c) at (529.85bp,18.0bp) [draw,ellipse,n] {\small{$\mat{M}^{(3)}$}};
  \node (M4_c) at (611.85bp,18.0bp) [draw,ellipse,n] {\small{$\mat{M}^{(4)}$}};
  \node (M1_c) at (693.85bp,18.0bp) [draw,ellipse,n] {\small{$\mat{M}^{(1)}$}};
  \node (M2_d) at (775.85bp,18.0bp) [draw,ellipse,n] {\small{$\mat{M}^{(2)}$}};
  \node (M3_d) at (858.85bp,18.0bp) [draw,ellipse,n] {\small{$\mat{M}^{(3)}$}};
  \node (M4_d) at (941.85bp,18.0bp) [draw,ellipse,n] {\small{$\mat{M}^{(4)}$}};
  \draw [->] (X) ..controls (443.74bp,215.37bp) and (309.46bp,186.18bp)  .. (M123);
  \draw [->] (X) ..controls (499.89bp,209.07bp) and (486.16bp,195.9bp)  .. (M124);
  \draw [->] (X) ..controls (538.55bp,208.4bp) and (544.34bp,198.02bp)  .. (M134);
  \draw [->] (X) ..controls (595.31bp,214.18bp) and (693.86bp,187.84bp)  .. (M234);
  \draw [->] (M123) ..controls (171.41bp,136.82bp) and (153.73bp,123.22bp)  .. (M12_a);
  \draw [->] (M123) ..controls (202.85bp,135.98bp) and (202.85bp,126.71bp)  .. (M23_a);
  \draw [->] (M124) ..controls (420.41bp,136.82bp) and (402.73bp,123.22bp)  .. (M14_b);
  \draw [->] (M124) ..controls (451.85bp,135.98bp) and (451.85bp,126.71bp)  .. (M12_b);
  \draw [->] (M134) ..controls (563.85bp,135.98bp) and (563.85bp,126.71bp)  .. (M34_c);
  \draw [->] (M134) ..controls (603.5bp,137.48bp) and (630.05bp,121.73bp)  .. (M14_c);
  \draw [->] (M234) ..controls (786.85bp,135.98bp) and (786.85bp,126.71bp)  .. (M23_d);
  \draw [->] (M234) ..controls (829.29bp,137.79bp) and (859.2bp,121.43bp)  .. (M34_d);
  \draw [->] (M12_a) ..controls (83.255bp,63.982bp) and (68.469bp,51.044bp)  .. (M1_a);
  \draw [->] (M12_a) ..controls (112.56bp,63.983bp) and (112.83bp,54.712bp)  .. (M2_a);
  \draw [->] (M23_a) ..controls (201.78bp,63.983bp) and (201.38bp,54.712bp)  .. (M3_a);
  \draw [->] (M14_b) ..controls (332.69bp,64.048bp) and (318.21bp,51.217bp)  .. (M4_b);
  \draw [->] (M14_b) ..controls (362.28bp,63.983bp) and (362.81bp,54.712bp)  .. (M1_b);
  \draw [->] (M12_b) ..controls (450.42bp,63.983bp) and (449.89bp,54.712bp)  .. (M2_b);
  \draw [->] (M34_c) ..controls (551.77bp,64.129bp) and (546.81bp,53.925bp)  .. (M3_c);
  \draw [->] (M34_c) ..controls (580.92bp,64.106bp) and (588.35bp,53.273bp)  .. (M4_c);
  \draw [->] (M14_c) ..controls (686.07bp,64.346bp) and (687.67bp,55.027bp)  .. (M1_c);
  \draw [->] (M23_d) ..controls (782.92bp,63.983bp) and (781.46bp,54.712bp)  .. (M2_d);
  \draw [->] (M23_d) ..controls (812.56bp,64.001bp) and (825.16bp,51.752bp)  .. (M3_d);
  \draw [->] (M34_d) ..controls (924.47bp,64.059bp) and (928.22bp,54.331bp)  .. (M4_d);
    \node[left=3em] at (M1_a)  (l3) {Level 3};
    \node at (l3 |- M12_a) (l2){Level 2};
    \node at (l3 |- M123) (l1) {Level 1};
\end{tikzpicture}
\caption{Multi-sweep dimension tree with $N=4$.
}
\label{fig:msdt}
\end{figure*}

The standard
dimension tree for CP-ALS constructs the tensor contraction paths based on a fixed amortization scheme for different ALS sweeps. It requires two first level TTM calculations, one for the MTTKRP of right-half modes and the other for the left-half modes, as is shown in Figure~\ref{subfig:dt}. 
However, cost of CP-ALS can be reduced further by amortizing first-level TTM contractions across sweeps.
Given an order $N$ tensor $\tsr{T}$, the first level TTM $\tsr{T}\times_i\mat{A}^{(i)}$ can be used for the MTTKRP of all the modes except $i$.
For example, with order $N=4$, the contraction between $\tsr{T}$ and $\mat{A}^{(4)}$ can be used for the construction of $\mat{M}^{(1)},\mat{M}^{(2)},\mat{M}^{(3)}$.
After this, the TTM with $\mat{A}^{(3)}$ can be used to compute not only the remaining term for this sweep, $\mat{M}^{(4)}$, but also $\mat{M}^{(1)}$ and $\mat{M}^{(2)}$ for the next sweep.
Given that each first-level TTM can be used to compute $N-1$ terms $\mat{M}^{(i)}$, we should be able to compute $N-1$ sweeps using $N$ such TTMs, as opposed to the $2(N-1)$ needed by a typical dimension tree.

Our multi-sweep dimension tree (MSDT) algorithm achieves this goal, and we illustrate the tree in Figure~\ref{fig:msdt}.
Each MSDT tree is responsible for the MTTKRP calculations of $N-1$ sweeps. It includes $N$ subtrees, and the root of $i$th subtree is the first level contraction $\tsr{T}\times_{N-i+1}\mat{A}^{(N-i+1)}$, which is used for the MTTKRP of all the $N-1$ modes except $N-i+1$, with the calculation order being $(N-i+1) + 1 \mod N \prec \cdots \prec (N-i+1) + N-1 \mod N$. 
Each subtree can be constructed with the traditional binary dimension tree.
MSDT achieves the computational cost of 
\[
2\frac{N}{N-1}s^NR + \bigo(s^{N-1}R)
\]
for each sweep, and the leading order cost is only $\frac{N}{2(N-1)}$ times that of the state-of-art dimension tree, thus speeding up the CP-ALS algorithm.

\section{Parallel Algorithms}
\label{sec:par-alg}

The parallel CP-ALS algorithm introduced in Section~\ref{subsec:par-als} can be easily combined with the MSDT algorithm, where only the DT routine in the local-MTTKRP calculations need to be replaced by the MSDT routine. The computational cost will be cheaper, and the horizontal communication cost will be the same for both algorithms.

\begin{algorithm}[]
    \caption{\textbf{Par-PP-CP-ALS-subroutine}}
\label{alg:par_pp}
\begin{algorithmic}[1]
\small
    \STATE{
    \textbf{Assume:} Local matrices $\mt{A}{i}{\proc{P}(x)}, d\mt{A}{i}{\proc{P}(x)}\text{  }\forall i\in\{1,\ldots,N\}$ prepared for local-MTTKRPs
    }
    \STATE{Call $\text{Local-PP-init}(\tsr{T}_{\proc{P}(x)}, \{\mt{A}{1}{\proc{P}(x)},\ldots,\mt{A}{N}{\proc{P}(x)}\})$
    to update all $\tsr{M}^{(i,n)}_{p,\proc{P}(x)},\mat{M}^{(n)}_{p,\proc{P}(x)}$
    \label{line:local_pp_init_step}}
    \WHILE{$\forall ~i \in \inti{1}{N}, \fnrm{d\mat{A}^{(i)}}<\epsilon\fnrm{\mat{A}^{(i)}}$}
      \FOR{\texttt{$n\in \inti{1}{N} $}}
        \FOR{\texttt{$i\in \inti{1}{N}, i\neq n $}}
            \STATE Update $\mat{U}_{\proc{P}(x)}^{(n,i)}$ with 
    
            $\mat{U}_{\proc{P}(x)}^{(n,i)}(a,k) \leftarrow \sum_{b}\tsr{M}_{p,\proc{P}(x)}^{(n,i)}(a,b,k) d\mat{A}_{\proc{P}(x)}^{(i)}(b,k)$\label{line:local_pp_update_M}
        \ENDFOR
        \STATE{
        $\Tilde{\mat{M}}_{\proc{P}(x)}^{(n)}
    \leftarrow \mat{M}_{p,\proc{P}(x)}^{(n)} + \sum_{i=1,i\neq n}^{N}\mat{U}_{\proc{P}(x)}^{(n,i)} $
         }
    \STATE{Update global $\Tilde{\mat{M}}^{(n)}$ based on $\Tilde{\mat{M}}_{\proc{P}(x)}^{(n)}$} \label{line:pp_update_global_M}
        \STATE Calculate $\mat{V}^{(n)}$ based on Equation~\eqref{eq:V_nij} \label{line:par-V}
        \STATE{
        $\Tilde{\mat{M}}^{(n)}
    \leftarrow\Tilde{\mat{M}}^{(n)}
    + \mat{V}^{(n)} $
         }
        \STATE{$\boldsymbol{\Gamma}^{(n)}\leftarrow\mat{S}^{(1)}\ast\cdots\ast \mat{S}^{(n-1)}\ast \mat{S}^{(n+1)}\ast\cdots\ast \mat{S}^{(N)} $}
        \STATE{$   \mat{A}^{(n)} \leftarrow \Tilde{\mat{M}}^{(n)}\boldsymbol{\Gamma}^{(n)}{}^{\dagger} $}
        \STATE{$d\mat{A}^{(n)} = \mat{A}^{(n)} -\mat{A}^{(n)}_p$, $   \mat{S}^{(n)} \leftarrow {\mat{A}^{(n)}}{}^T\mat{A}^{(n)} $}
    	\STATE Prepare local factors $\mt{A}{n}{\proc{P}(x)}, d\mt{A}{n}{\proc{P}(x)}$ from $\mat{A}^{(n)}, d\mat{A}^{(n)}$
      \ENDFOR
    \ENDWHILE
\end{algorithmic}
\end{algorithm}

\begin{table*}[]
\caption{Comparison of the leading order sequential computational cost, leading order local computational cost, asymptotic communication cost and the leading order auxiliary memory necessary on each processor for the MTTKRP calculation. PP-init-ref, PP-approx-ref denote the PP initialization and approximated kernels implemented in the reference~\cite{ma2018accelerating}.} 
\label{table:compare}
\centering
\begin{adjustbox}{width=.995\textwidth}
\begin{tabular}{|c|c|c|c|p{0.32\textwidth}|c|}
  \hline
 & Sequential computational cost
 & Local computational cost
 & Auxiliary memory
 & \centering Horizontal communication cost
 & Vertical communication cost
 \\ 
  \hline
DT & $4s^NR$  & $4s^NR/P$ 
& $ (s^N/P)^{1/2}R$
&\centering$\bigo(N\log(P)\cdot\alpha + NsR/P^{\frac{1}{N}}\cdot\beta)$ 
& $\bigo((s^N/P + (s^N/P)^{1/2}R)\cdot \nu)$ \\ 
  \hline
  MSDT & $\frac{2N}{N-1}s^NR$ & $\frac{2N}{N-1}s^NR/P$ 
    & $(s^N/P)^{\frac{N-1}{N}}R$
  & \centering$\bigo(N\log(P)\cdot\alpha + NsR/P^{\frac{1}{N}}\cdot\beta)$ 
& $\bigo((s^N/P + (s^N/P)^{\frac{N-1}{N}}R)\cdot \nu)$
 \\ 
  \hline
PP-init & $4s^NR$ & $4s^NR/P$ 
& $(s^N/P)^{\frac{N-1}{N}}R$ 
& \centering / 
& $\bigo((s^N/P + (s^N/P)^{\frac{N-1}{N}}R)\cdot \nu)$
\\
  \hline
PP-init-ref & $4s^NR$ & $4s^NR/P$ 
& $s^{N-1}R/P$
& \centering$\bigo(N\log(P)\cdot\alpha + N(s^{N}R/P)^{2/3} \cdot \beta)$ or 
$\bigo(N\log(P)\cdot\alpha + N(s^N/p)^{\frac{N-1}{N}}R \cdot \beta)$ 
& $\bigo((s^N/P + (s^N/P)^{\frac{N-1}{N}}R)\cdot \nu)$
 \\
  \hline
PP-approx & $2N^2(s^2R + R^2) $ & 
$2N^2(s^2R/P^{\frac{2}{N}} + R^2/P) 
 $ 
  & $N^2s^2R/P^{\frac{2}{N}} + NR^2/P$
 & \centering$\bigo(N\log(P)\cdot\alpha + NsR/P^{\frac{1}{N}}\cdot\beta)$ 
 & $\bigo(N^2(s^2R/P^{\frac{2}{N}} + R^2/P) \cdot \nu)$
  \\
  \hline
PP-approx-ref & $2N^2(s^2R + R^2) $ & 
$2N^2(s^2R/P + R^2/P) $ 
 & $N^2s^2R/P + NR^2/P$
& \centering$\bigo(N^2\log(P)\cdot\alpha + N^2sR/P\cdot\beta)$
 & $\bigo(N^2(s^2R/P + R^2/P) \cdot \nu)$
  \\
  \hline
\end{tabular}
\end{adjustbox}
\end{table*} 

We detail the subroutine of the parallel PP algorithm in Algorithm~\ref{alg:par_pp}. In the algorithm, we show the parallel version of the PP initialization step and the approximated step (lines~\ref{line:pp_init_step}-\ref{line:pp_endif} in Algorithm~\ref{alg:cp_als_pp}). In Algorithm~\ref{alg:par_pp}, the core idea is to perform all the contractions in the PP initialization steps (line~\ref{line:local_pp_init_step}), and first-order corrections in the PP approximated steps (line~\ref{line:local_pp_update_M}) locally, similar to the local-MTTKRP routine. 
After the calculations of the local first-order MTTKRP corrections, we use the Reduce-Scatter collective routine to update the global $\Tilde{\mat{M}}^{(n)}$ (line~\ref{line:pp_update_global_M}). 
In addition, the second-order correction (line~\ref{line:par-V}) only involves Hadamard products and a small matrix multiplication, and we calculate that in parallel.

We show the comparison of the computational cost, communication cost and the auxiliary memory needed among DT, MSDT and PP in Table~\ref{table:compare}. The leading order computational cost of 
MSDT is a factor of $\frac{2(N-1)}{N}$ smaller than the cost of DT.
The PP initialization step has the same leading order cost as DT, and the PP approximated step reduces the local computational cost to $\bigo(N^2(s^2R/P^{\frac{2}{N}} + R^2/P) )$. 

As to the auxiliary memory usage, both MSDT and PP require more memory compared to DT, to achieve the optimal computational cost shown in the table. Both MSDT and PP can use less memory through combining several upper level contractions.
If the upper $l \leq N - 2$ levels of contractions are combined, both PP and MSDT would require 
$(s^N/P)^{\frac{N-l}{N}}R$ local  auxiliary memory. However, the local computational cost of the PP initialization step would increase to $(l+2)(l+1)s^{N}R/P$, and the local computational cost of MSDT would increase to $\frac{2N}{N-l}s^NR/P$.

DT, MSDT and PP all have the same asymptotic horizontal communication cost. In addition, our current PP implementation is more efficient in horizontal communication than that in~\cite{ma2018accelerating}, which regards the PP initialization step as a general matrix multiplication. It will either keep the input tensor in place, perform local multiplications and afterwards perform a reduction on the output tensor when $R$ is small, or perform a general 3D parallel matrix multiplication when $R$ is high. For both cases, the communication cost is high. 

As to the vertical communication cost, both MSDT and the PP initialization step have higher costs compared to DT, on account of the larger intermediates formed in the dimension trees. For high order cases, one can combine the upper $l$ levels of contractions to achieve a trade-off between the computational cost and the vertical communication cost. The vertical communication cost of the PP approximated step is greater than the local computational cost, inferring that the PP approximated step is vertical communication bound. Our performance evaluation in Section~\ref{subsec:bench} 
confirms this.

Note that tensor transposes are not necessary for the contractions in DT but are necessary in the PP initialization step when the order is greater than 3 (for the PP operators calculations), and in MSDT to contract the input tensor with an middle mode factor matrix, e.g. $\mat{A}^{(2)}$. 
Performing tensor transposes will incur a larger leading order constant in the vertical communication cost. 
The transposes in the PP initialization step will affect the PP performance, as we will show in Section~\ref{sec:results}. The transposes in MSDT can be avoided if transposes of the input tensor is stored. For both order 3 and order 4 tensors, one copy of the transposed input tensor is necessary. Our implementations use this method to avoid transposes in MSDT.

\section{Experimental Results}
\label{sec:results}

\newdimen\figrasterwd
\figrasterwd\textwidth
\begin{figure*}
  \centering
  \parbox{\figrasterwd}{
    \parbox{.64\figrasterwd}{%
\subfloat[$N=3,s_{\text{local}}=400, R=400$]{
\includegraphics[width=.33\textwidth]{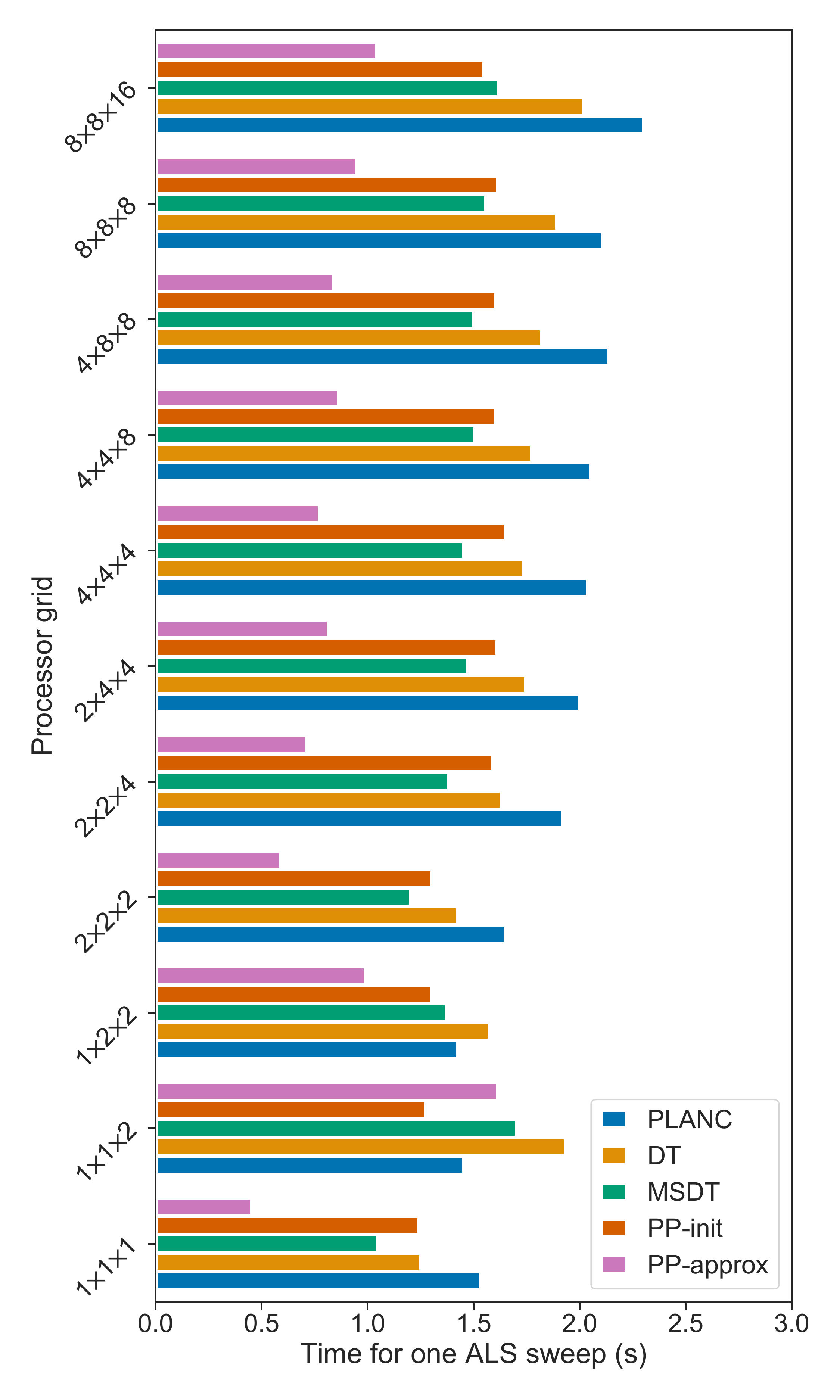}\label{subfig:weakscale-3d}
}
\subfloat[$N=4,s_{\text{local}}=75, R=200$]{
\includegraphics[width=.33\textwidth]{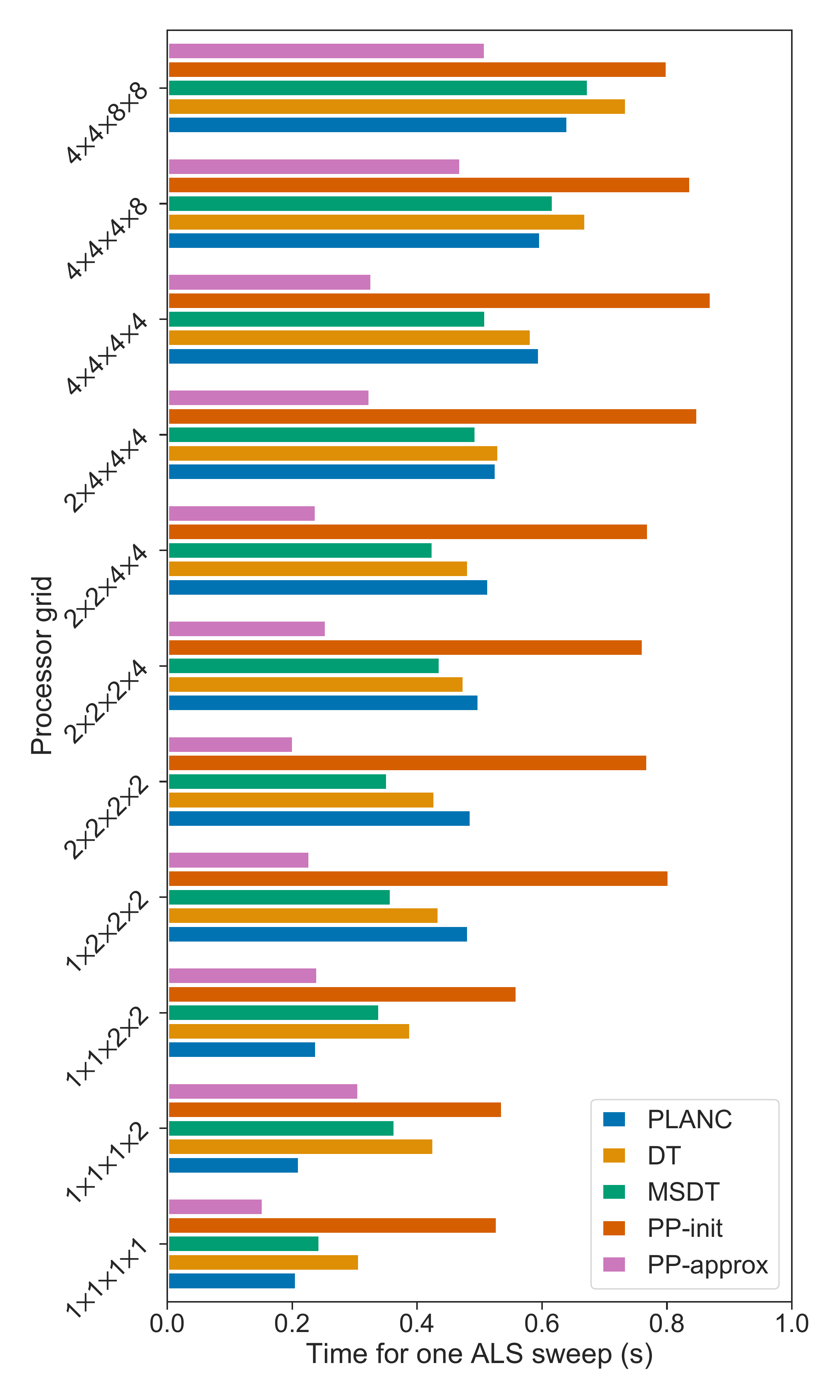}\label{subfig:weakscale-4d}
}
    }
    \hskip1em
    \parbox{.32\figrasterwd}{%
\subfloat[$N=3,$ grid $=2\times 4\times 4$]{
\includegraphics[width=.33\textwidth]{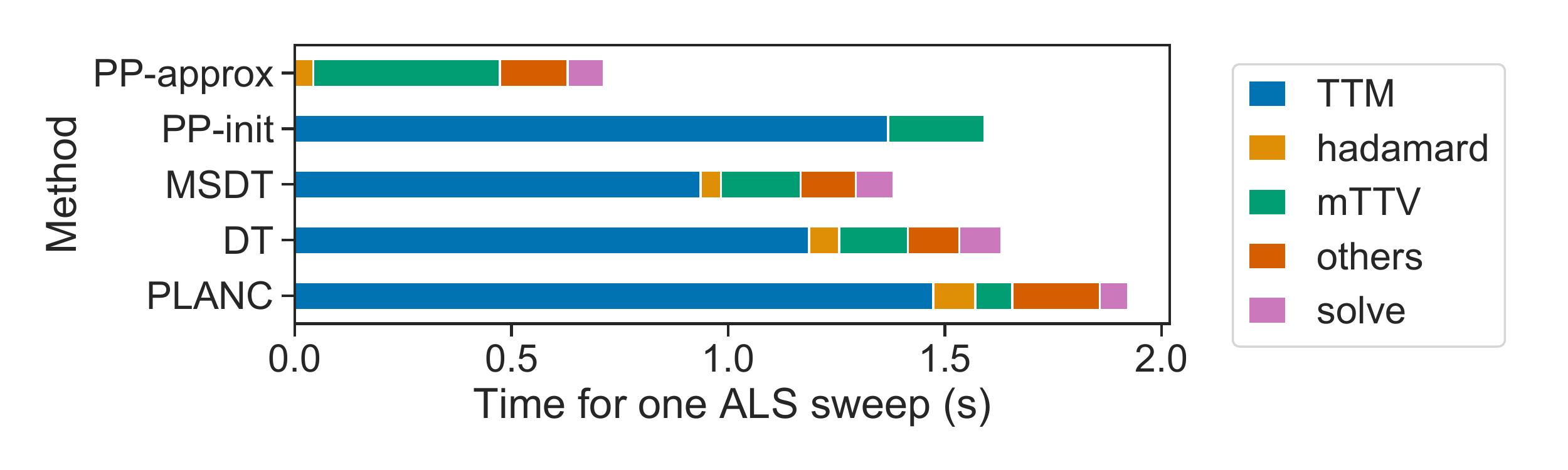}\label{subfig:timebreak-3d1}
}
\vspace{-0mm}
\subfloat[$N=3,$ grid $=8\times 8\times 8$]{
\includegraphics[width=.33\textwidth]{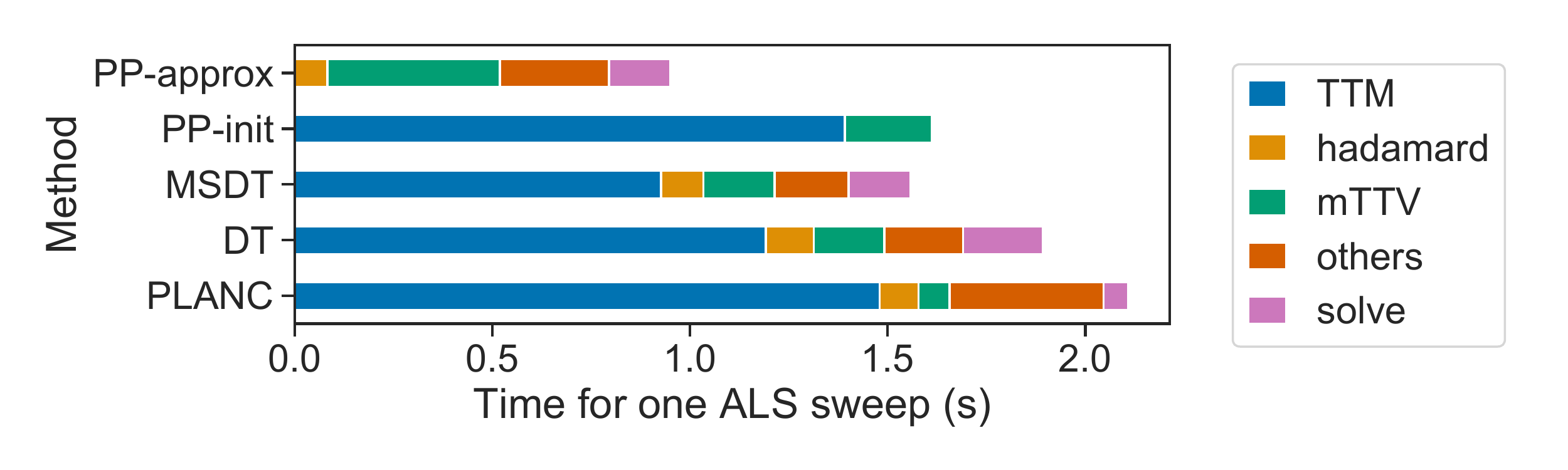}\label{subfig:timebreak-3d2}
}
\vspace{-0mm}
\subfloat[$N=4,$ grid $=2\times 2\times 2\times 2$]{
\includegraphics[width=.33\textwidth]{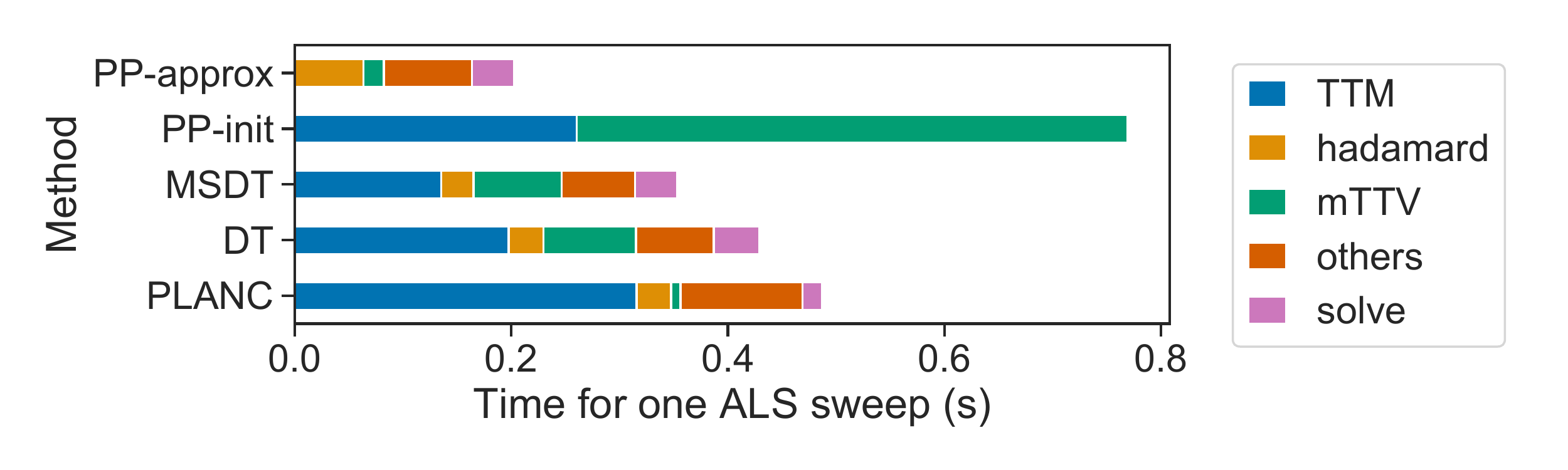}\label{subfig:timebreak-4d1}
}
\vspace{-0mm}
\subfloat[$N=4,$ grid $=4\times 4\times 4\times 4$]{
\includegraphics[width=.33\textwidth]{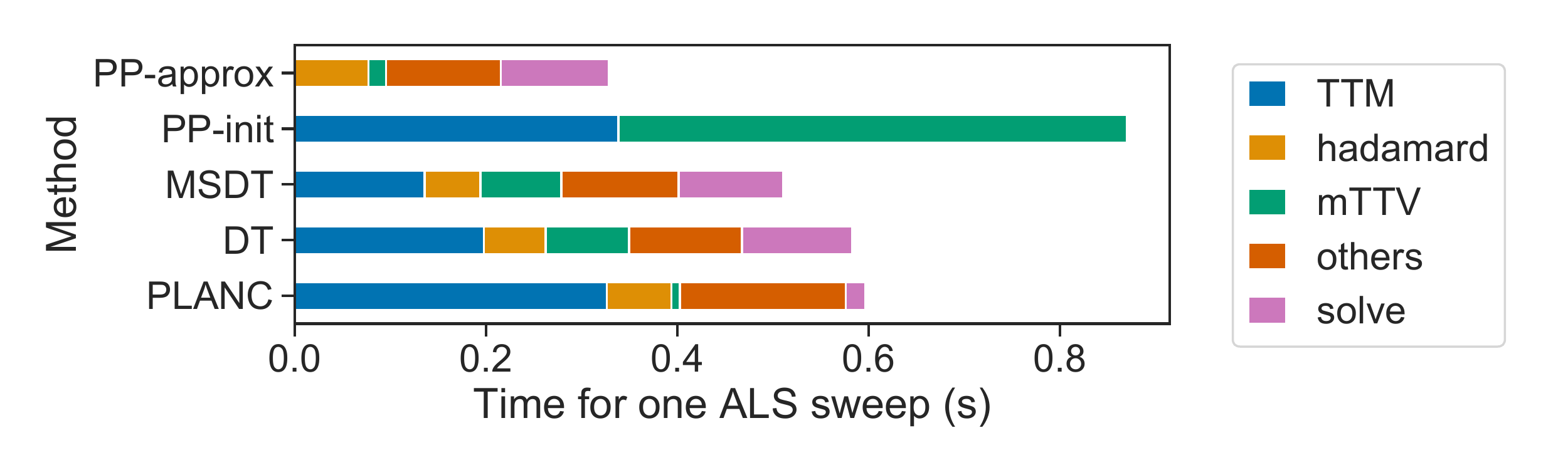}\label{subfig:timebreak-4d2}
}
}}
\caption{Benchmark results for order 3 and 4 tensors. The reported time is the mean time across 5 sweeps. \textbf{(a)(b)} Weak scaling of synthetic tensors. For each plot, the rank $R$ is fixed, and the local tensor size on each processor is fixed with dimension size $s_{\text{local}}$. \textbf{(c)(d)(e)(f)} Time breakdown under specific processor grid configurations. Each per-sweep time is broken into 5 categories: TTM, mTTV, solve (linear systems solves), hadamard (the Hadamard products, which appear in Equations~\ref{eq:gram},\ref{eq:V_nij}), and others.
The tensor sizes and CP ranks of (c)(d) are the same as in (a), and of (e)(f) are the same as in (b).}
\label{fig:benchmark}
\end{figure*}

\begin{table*}[]
\caption{Comparison of the per-sweep MTTKRP calculation time between our PP initialization step (PP-init) and the PP approximation step (PP-approx) kernels to the ones (PP-init-ref, PP-approx-ref) implemented in the reference~\cite{ma2018accelerating}. The tensor size and CP ranks under each processor grid configuration is the same as in Figure~\ref{subfig:weakscale-3d},\ref{subfig:weakscale-4d}. 
} 
\label{table:compare_with_ma2018}
\centering
\begin{adjustbox}{width=.99\textwidth}
\begin{tabular}{|c|c|c|c|c|c|c|c|c|}
\hline
  Processor grid
 & $2\times 4 \times 4$ (3D)
 & $4\times 4 \times 4$ (3D)
 & $4\times 4 \times 8$ (3D)
 & $4\times 8 \times 8$ (3D)
 & $2\times 2 \times 2 \times 4$ (4D)
 & $2 \times 2\times 4 \times 4$ (4D)
 & $2\times 4\times 4 \times 4$ (4D)
 & $4\times 4\times 4 \times 4$ (4D)
 \\ 
  \hline
PP-init & 1.6105 & 1.6535 & 1.6045 & 1.6060 & 0.7627 & 0.7713 &  0.85 & 0.8715 \\ 
  \hline
PP-init-ref & 12.9300 & 12.1920 & 11.9075 & 11.3710 & 19.8695 & 19.6200 & 16.018 & 13.3695 \\
  \hline
PP-approx & 0.4579 & 0.4509 & 0.4410 & 0.4433 & 0.0541 & 0.0526  & 0.0553 & 0.0533 \\ 
  \hline
PP-approx-ref & 6.7128 & 5.7927 & 5.3655 & 4.5682 & 0.3540 & 0.2916 & 0.2757 & 0.2887 \\ 
  \hline
\end{tabular}
\end{adjustbox}
\end{table*}

\subsection{Implementations, Platforms and Tensors}

We implement parallel DT, MSDT and PP algorithms with Cyclops Tensor Framework (v1.5.5)~\cite{solomonik2014massively}, which is a distributed-memory library for matrix/tensor contractions that uses MPI for interprocessor communication and OpenMP for threading. On each processor, we use Intel compilers and the MKL library for threaded BLAS routines, including batched BLAS routines, which are efficient for mTTV arising in MTTKRP, and employ the HPTT library~\cite{springer2017hptt} for high-performance tensor transpositions.
We also use a wrapper provded by Cyclops for ScaLAPACK~\cite{Dongarra:1997:SUG:265932} routines to solve symmetric positive definite linear systems of equations\footnote{All of our code is available at \url{https://github.com/LinjianMa/parallel-pp}.}. All storage and computations assume the tensors are dense. Our PP algorithm calls the MSDT subroutine for the regular ALS sweeps (line~\ref{line:regular_als} in Algorithm~\ref{alg:cp_als_pp}) to improve the performance. 

The experimental results are collected on the Stampede2 supercomputer located at the University of Texas at Austin.
Experiments are performed on the Knight's Landing (KNL) nodes, each of which consists of 68 cores, 68 threads, 96 GB of DDR RAM, and 16 GB of MCDRAM.
These nodes are connected via a 100 Gb/sec fat-tree Omni-Path interconnect.

We compare the performance of different algorithms on both synthetic tensors and application datasets. 
The application datasets we considered include publicly available tensor datasets as well as tensors of interest for quantum chemistry calculations.
These tensors enable us to demonstrate the effectiveness of our algorithms on practical problems. 
We use the following four tensors to test the performance. For all the experiments, we use 16 processors on each KNL node, and each processor uses 4 threads. 
\begin{enumerate}[leftmargin=*]
    \item \label{tsr:col} \textbf{Tensors with given collinearity}~\cite{battaglino2017practical}. We create tensors based on randomly-generated factor matrices $\mat{A}^{(n)}$, where $n\in\{1,\ldots,N\}$. Each factor matrix $\mat{A}^{(n)} \in \mathbb{R}^{s\times R}$ is randomly generated so that the columns have collinearity defined based on a scalar $C$ (selected randomly from a given interval $[a,b)$),
$$
\frac{\langle \mat{a}_i^{(n)}, \mat{a}_j^{(n)}\rangle}{\vnrm{\mat{a}_i^{(n)}}\vnrm{\mat{a}_j^{(n)}}}
=C, \quad \forall i,j \in \{1, \ldots, R\}, i\neq j. 
$$
The generated tensor has dimension size $s$ in each mode, and its CP rank is bounded by $R$.
Higher collinearity corresponds to greater overlap between columns within each factor matrix, which makes the convergence of the rank-$R$ CP-ALS procedure slower~\cite{rajih2008enhanced}. 
We experiment on tensors with dimensions $1600\times 1600\times 1600$. We run the experiments on 64 processors, and the processor grid has dimension $4 \times 4 \times 4$.
\item \label{tsr:quantum}\textbf{Quantum chemistry tensor.} We also test on the density fitting intermediate tensor arising in quantum chemistry, which is the Cholesky factor of the two-electron integral tensor~\cite{hohenstein2012tensor,hummel2017low}. For the order 4 two-electron integral tensor $\tsr{T}$, its Cholesky factor is an order 3 tensor $\tsr{D}$, with their element-wise relation shown as follows:
\[
\tsr{T}(a,b,c,d) = \sum_{e=1}^{E}\tsr{D}(a,b,e)\tsr{D}(c,d,e),
\]
where $E$ is the third mode dimension size of $\tsr{D}$.
CP decomposition can be performed on $\tsr{D}$ to compress the intermediate and can accelerate the post Hartree-Fork calculations~\cite{hohenstein2012communication}. We generate the density fitting tensor via the PySCF library~\cite{sun2018pyscf}, which represents the compressed restricted Hartree-Fock wave function of an 40 water molecule chain system with a STO-3G basis set. The generated tensor has size $4520\times 280\times 280$. We run the experiments on 32 processors, and the processor grid has dimension $8 \times 2 \times 2$.
\item  \label{tsr:coil} \textbf{COIL dataset}. COIL-100 is an image-recognition dataset that contains images of objects in different poses~\cite{nenecolumbia}. It has been used previously as a tensor decomposition benchmark~\cite{battaglino2017practical,zhou2014decomposition,ma2018accelerating}.
Transferring the data into the tensor format, the generated tensor has size $128\times 128\times 3\times 7200$. We run the experiments on 16 processors, and the processor grid has dimension $2 \times 4 \times 1 \times 2$.
\item \label{tsr:timelapse}  \textbf{Time-Lapse hyperspectral radiance images}. We consider the 3D hyperspectral imaging dataset called “Souto wood pile”~\cite{nascimento2016spatial}. The dataset is usually used on the benchmark of nonnegative tensor decomposition~\cite{liavas2017nesterov,ballard2018parallel}. The hyperspectral data consists of a tensor with dimensions $1024\times1344\times33\times 9$. We run the experiments on 16 processors, and the processor grid has dimension $4 \times 4 \times 1 \times 1$.
\end{enumerate}

\begin{figure}[]   
\centering
\includegraphics[width=0.43\textwidth, keepaspectratio]{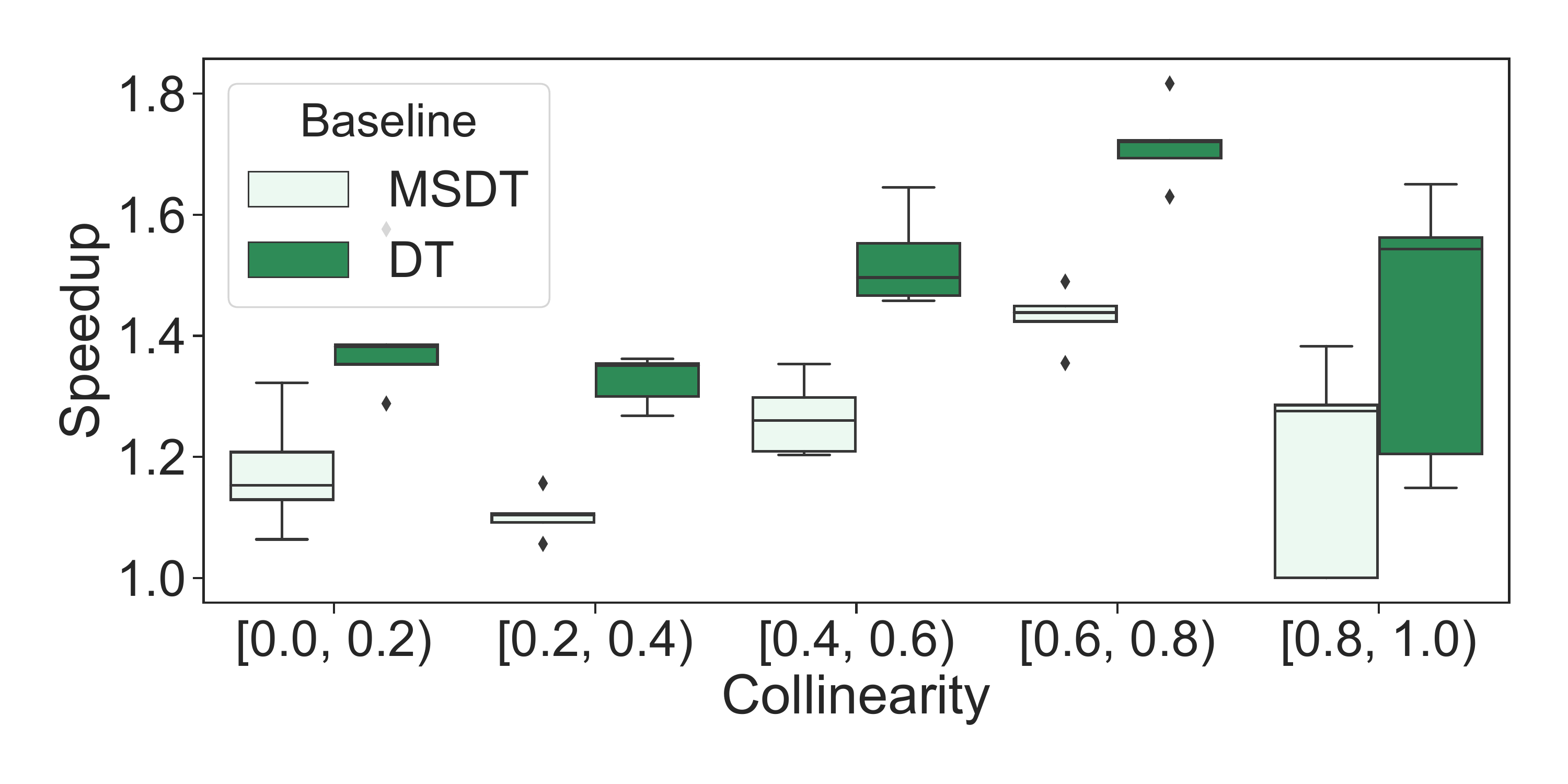}
\caption{Relation between PP speed-ups and input collinearity for order 3 tensors. The dimension size $s=1600$, the rank $R=400$, and the experiments run on the $4\times 4\times 4$ processor grid. The PP tolerance is set as 0.2. Each box is based on 5 experiments with different random seeds and shows the 25th-75th quartiles. The median is indicated by a horizontal line inside the box and outliers are displayed as dots.
}
\label{fig:box}
\end{figure}
\begin{table}[]
\caption{
Detailed statistics of the results shown in Figure~\ref{fig:box}. From left to right: the tensor configuration (Col stands for collinearity), number of exact ALS sweeps, number of PP initialization sweeps and number of PP approximated sweeps. All the data are the average statistics from 5 experiments.
} 
\label{table:randomstat}
\centering
\begin{adjustbox}{width=.42\textwidth}
\begin{tabular}{|c|c|c|c|}
  \hline
  Configuration
 & Num-ALS
 & Num-PP-init
 & Num-PP-approx
 \\ 
  \hline
Col$\in[0.0, 0.2)$ & 21.2 & 1.4 & 14.4  \\ 
  \hline
Col$\in[0.2, 0.4)$ & 50.8 & 11.6 & 39.6 \\ 
  \hline
Col$\in[0.4, 0.6)$ & 65.6 & 29.0 & 169.6 \\ 
  \hline
Col$\in[0.6, 0.8)$ & 34.8 & 17.4 & 180.2 \\ 
  \hline
Col$\in[0.8, 1.0)$ & 10.2 & 3.4 & 22.8 \\ 
  \hline
\end{tabular}
\end{adjustbox}
\end{table}

\subsection{Benchmarks}
\label{subsec:bench}

We perform a parallel scaling analysis to compare
the per-ALS sweep simulation time for DT, MSDT, the PP initialization step and the PP approximated step, and the results are presented in Figure~\ref{fig:benchmark}. We also show the simulation time using PLANC~\cite{eswar2019planc} for reference, which contains the state-of-art parallel DT implementations. The benchmarks are performed on both order 3 and order 4 tensors. 

We show the order 3 weak scaling results in Figure~\ref{subfig:weakscale-3d}. 
As can be seen, the performance of our DT implementation is comparable, and slightly better than PLANC. MSDT performs consistently better than DT. Under the largest processor grid configuration, MSDT is 1.25X better than DT. In addition, the time spent on each PP initialization step is consistently less than each DT sweep, and the PP approximated step achieves a speed-up of $1.94$X compared to DT under the largest processor grid, showing the good scalability of our parallel PP algorithm. We show the detailed time breakdown under the grid configurations $2\times 4 \times 4$, $8\times 8 \times 8$ in Figure~\ref{subfig:timebreak-3d1},\ref{subfig:timebreak-3d2}. The TTM is the major bottleneck for all the kernels except the PP approximated step, which is bounded by the mTTV kernel. Note that 
as is discussed in Section~\ref{sec:par-alg}, the mTTV kernel is vertical communication (memory bandwidth) bound, resulting in less experimental speed-ups despite the large speed-ups in terms of the flop counts.

We show the order 4 weak scaling results in Figure~\ref{subfig:weakscale-4d}.
Considering that the PP initialization step for the order 4 tensors involves several tensor transposes, we use 8 processors per KNL node and 8 threads per processor for the benchmark, so that the transposes can be accelerated with  a relatively large number of
threads on each processor. 
The performance of our DT implementation is comparable to PLANC for most of the processor grid configurations. When the number of processors used is large, our DT implementation is slightly slower than PLANC. This can be explained by the slow global linear system solves shown in Figure~\ref{subfig:timebreak-4d2}.
Since the CP rank is relatively small, using too many processors results in over-parallelization and the horizontal communication cost increases. MSDT performs consistently better than DT, and achieves a speed-up of 1.10X under the largest processor grid configuration. In addition, the PP initialization step is much slower than the DT sweep, which is due to several tensor transposes in the mTTV kernels. It also accounts for the slow mTTV kernels shown in Figure~\ref{subfig:timebreak-4d1},\ref{subfig:timebreak-4d2}. The PP approximated step achieves a speed-up of $1.44$X compared to DT under the largest processor grid. Overall, the speed-up of PP is less on order 4 compared to order 3 tensors. 

We also show the per-sweep MTTKRP calculation time comparison between our parallel PP implementation and that in~\cite{ma2018accelerating} in Table~\ref{table:compare_with_ma2018}. As can be seen, the current implementation is more efficient for both kernels under all the processor grid configurations. The local dimension tree calculations greatly decrease the horizontal communication cost.

\subsection{Performance Comparison}

\begin{figure}[]   
\label{fig:real}
\centering

\subfloat[Col$\in[0.6, 0.8)$, $R=400$]{\includegraphics[width=0.24\textwidth, keepaspectratio]{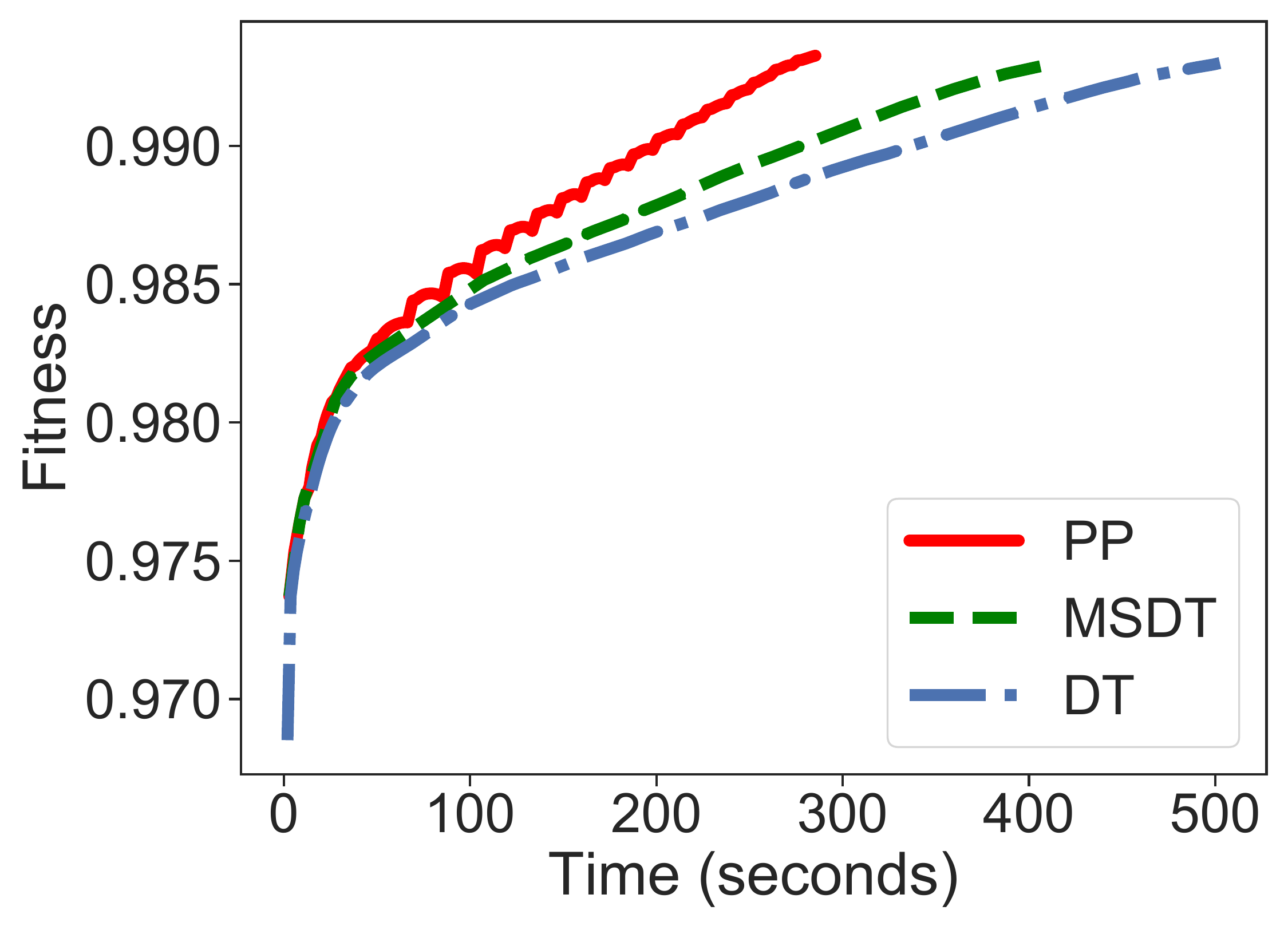}\label{detail:random}}
\subfloat[Chemistry, $R=300$]{\includegraphics[width=0.24\textwidth, keepaspectratio]{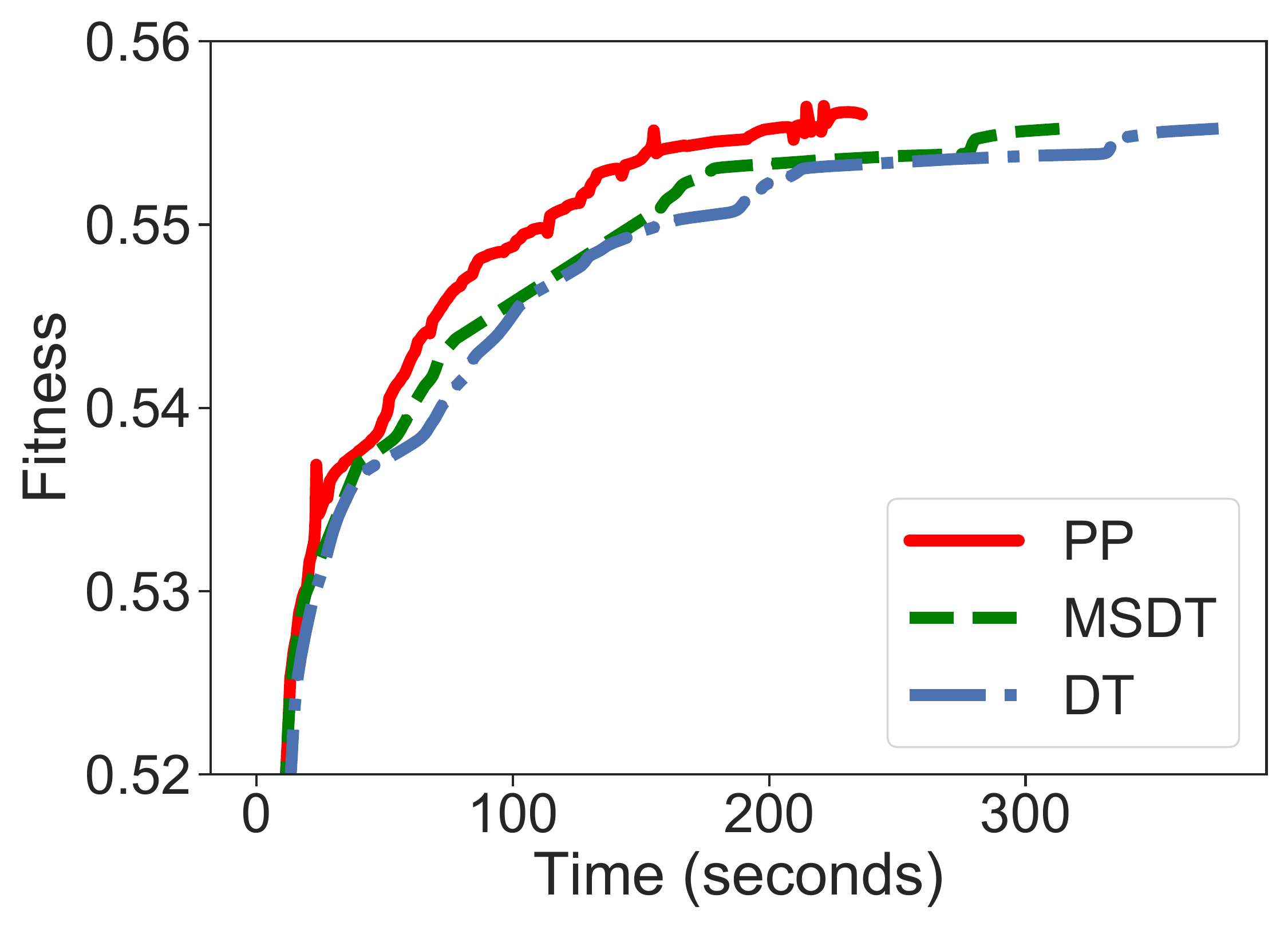}\label{detail:chem_r300}}

\subfloat[Chemistry, $R=600$]{\includegraphics[width=0.24\textwidth, keepaspectratio]{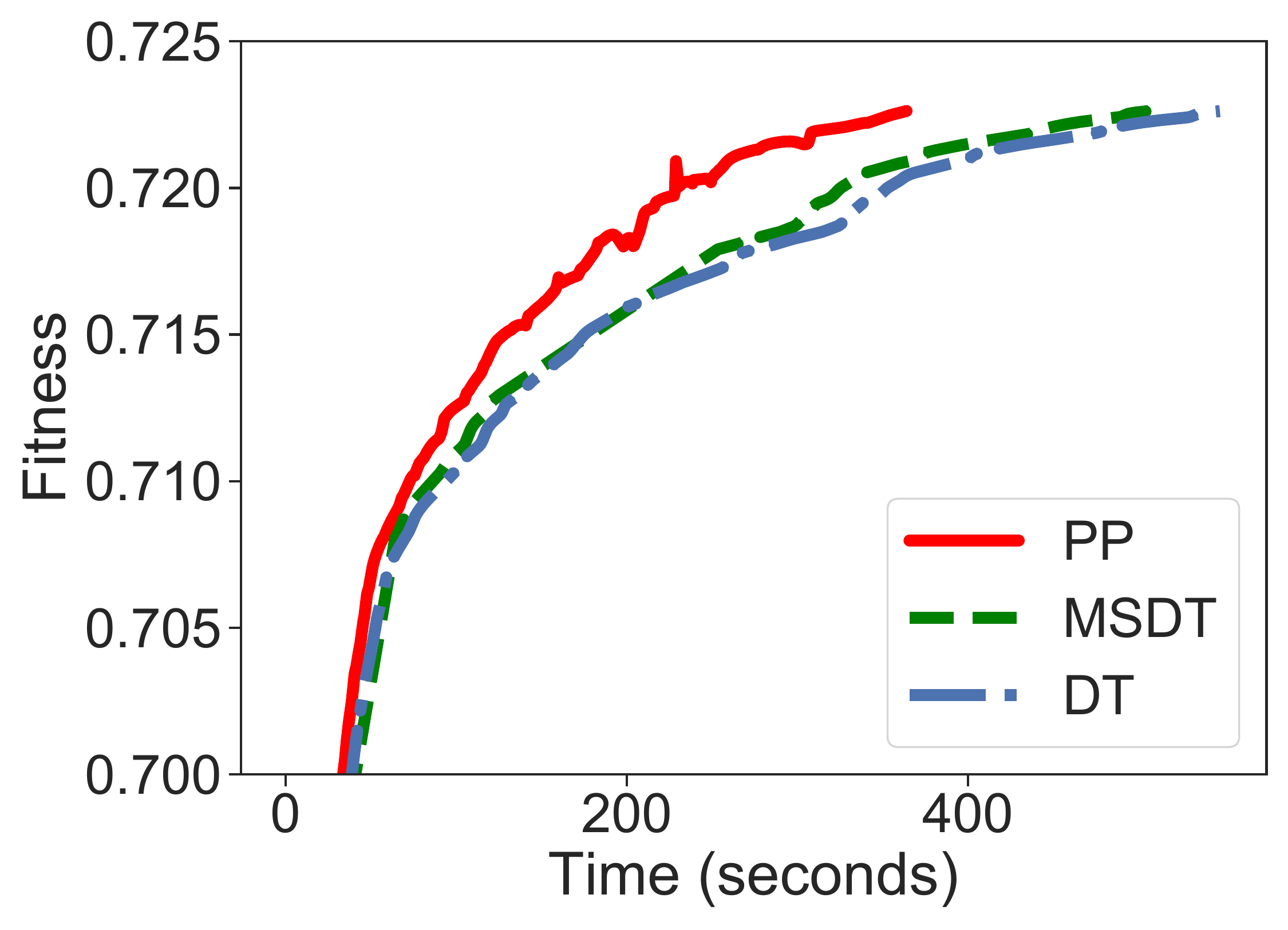}\label{detail:chem_r600}}
\subfloat[Chemistry, $R=1000$]{\includegraphics[width=0.24\textwidth, keepaspectratio]{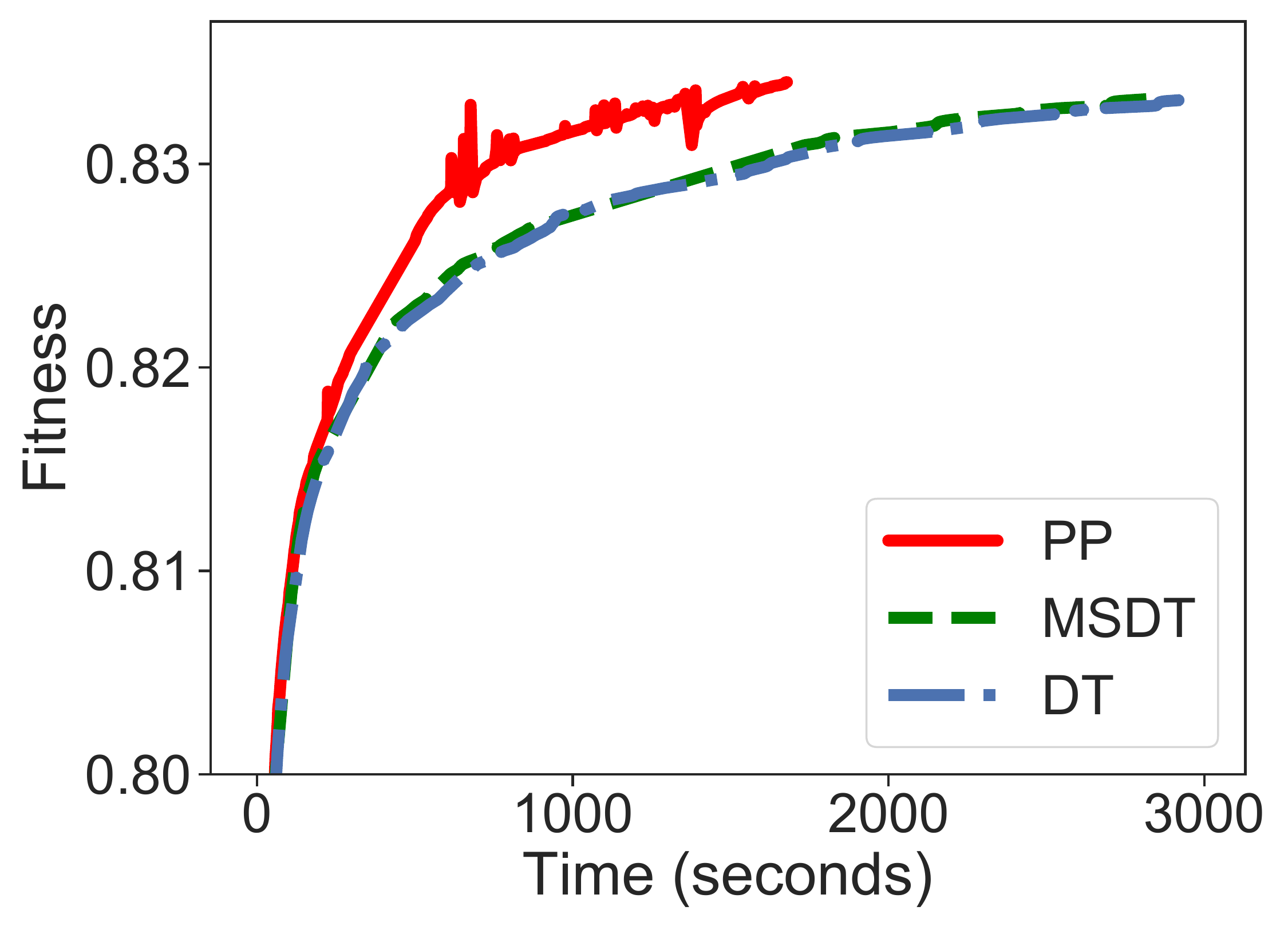}\label{detail:chem_r1000}}

\subfloat[Coil dataset, $R=20$]{\includegraphics[width=0.24\textwidth, keepaspectratio]{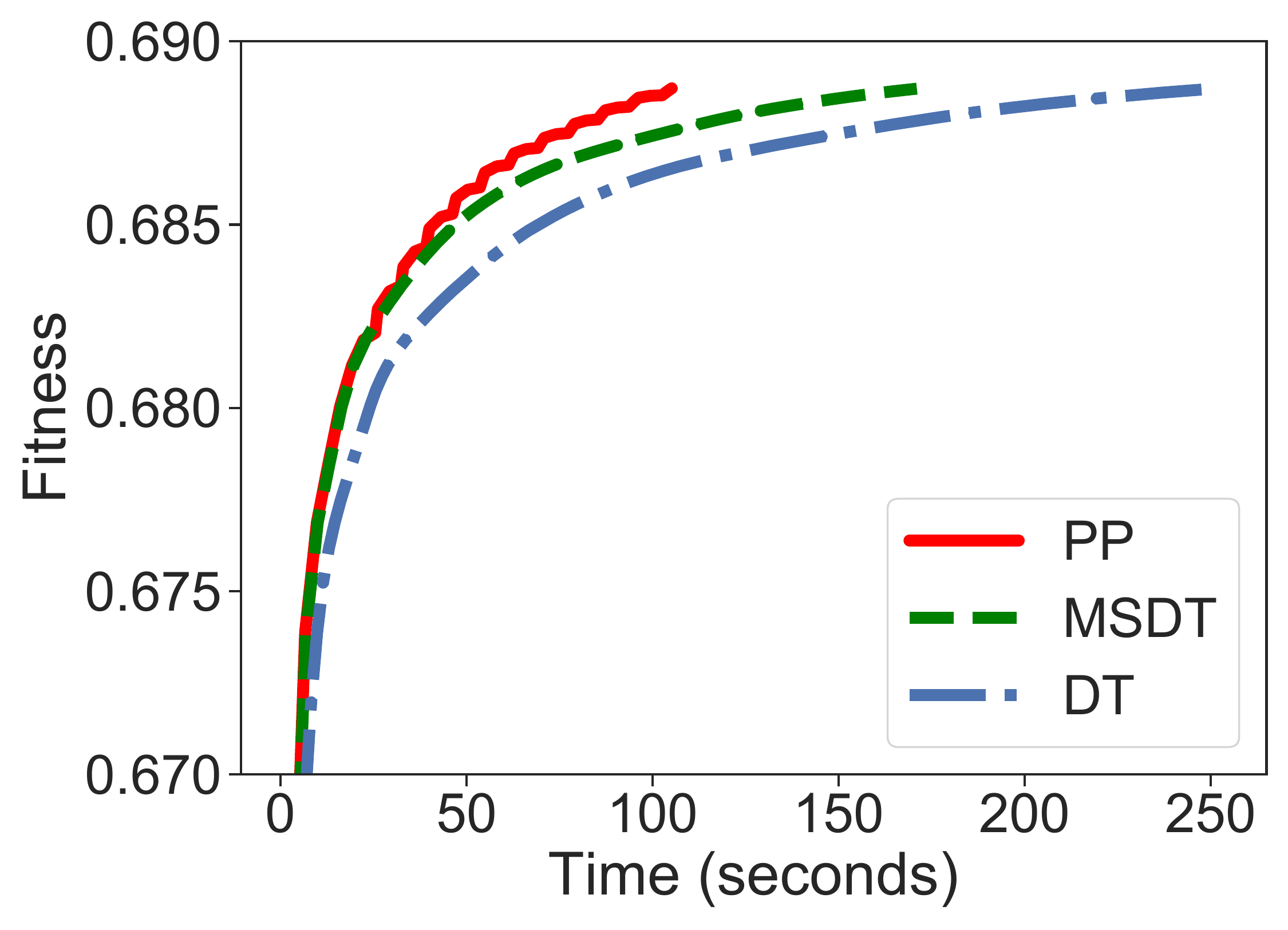}\label{detail:coil}}
\subfloat[Time-lapse dataset, $R=50$]{\includegraphics[width=0.24\textwidth, keepaspectratio]{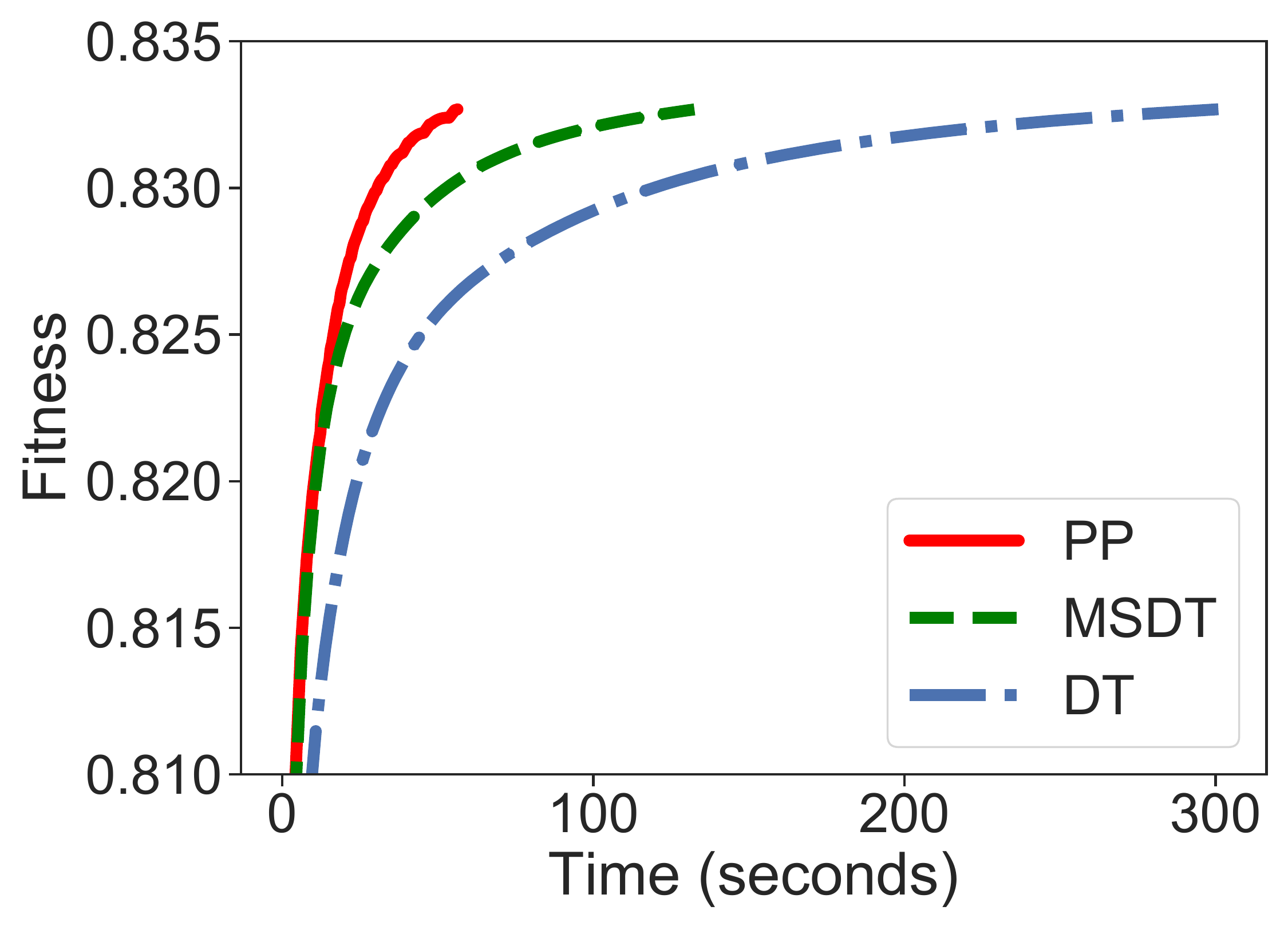}\label{detail:timelapse}}

\caption[]{
Comparison of PP, MSDT, DT on different tensors. 
}
\label{fig:detail_comparison}
\end{figure}

\begin{table}[]
\caption{
Detailed statistics of the results shown in Figure~\ref{fig:detail_comparison}. From left to right: the tensor type, number of ALS sweeps until PP experiments are finished, number of PP initialization sweeps, number of PP approximated sweeps, the average time of each ALS sweep, the average time of each PP initialization sweep, and the average time of each PP approximated sweep.
} 
\label{table:detail-run}
\centering
\begin{adjustbox}{width=.49\textwidth}
\begin{tabular}{|c|c|c|c|c|c|c|}
\hline
  Tensor
 & N-ALS
 & N-PP-init
 & N-PP-approx
 & T-ALS
 & T-PP-init
 & T-PP-approx
 \\ 
  \hline
Figure~\ref{detail:chem_r300} & 96 & 39 & 477 & 0.5064 & 0.2335 & 0.3443 \\ 
  \hline
Figure~\ref{detail:chem_r600} & 76 & 29 & 468 & 0.8177 & 0.3801  & 0.5839 \\ 
  \hline
Figure~\ref{detail:chem_r1000} & 216  & 63 & 1129 & 1.4307 & 0.795 & 1.1095 \\ 
  \hline
Figure~\ref{detail:coil} & 34  &  11 & 115 & 2.3907 & 4.2253 & 0.1785 \\ 
  \hline
Figure~\ref{detail:timelapse} & 30 & 10 & 155 & 0.6703 & 0.3631 & 0.1554 \\ 
  \hline
\end{tabular}
\end{adjustbox}
\end{table}

We test algorithms on both the synthetic tensors (Tensor~\ref{tsr:col}) and real datasets (Tensor~\ref{tsr:quantum},\ref{tsr:coil},\ref{tsr:timelapse}). We use the metrics relative residual, $r$ (defined in Equation~\eqref{eq:residual-def}), and fitness, $f=1-r$, to evaluate the convergence progress of the decomposition.  Figures~\ref{fig:box} shows the speed-up distribution with different exact factor matrices collinearity. We stop the algorithm when the stopping tolerance (defined as the fitness difference between two neighboring sweeps) reaches $10^{-5}$, or the maximum number of iterations (300) is reached.
It can be seen that PP achieves up to 1.8X speed-up compared to DT. In addition, PP tends to have higher speed-ups when the collinearity is within $[0.4, 0.8)$. This is because tensors within above collinearity range will converge in more sweeps, and more PP approximated sweeps are activated, as can be seen in Table~\ref{table:randomstat}. 
When the collinearity is in the range of $[0.0, 0.4)$ and $[0.8, 1.0)$, the experiments stop in small number of sweeps, which results in less benefit of PP.
In addition, we show the fitness-time relation for one experiment with the collinearity$\in[0.6, 0.8)$ in Figure~\ref{detail:random}. As can be seen, the fitness increases monotonically, indicating that PP controls the approximation error well. 
The results show that substantial performance gains on large synthetic tensors can be achieved for PP under the parallel execution, which is a supplement to the results in \cite{ma2018accelerating}, which shows that PP can speed-up CP decomposition on small (each tensor has size $400\times 400\times 400$) synthetic tensors.

We also compare the performance on application tensors. The PP tolerance is set as 0.1 for these tensors. Figure~\ref{detail:chem_r300},\ref{detail:chem_r600},\ref{detail:chem_r1000} show the results on the quantum chemistry tensor with different CP ranks.
The detailed statistics are shown in Table~\ref{table:detail-run}. 
As is shown in the figures, for all the variants of the CP ranks, PP performs
better than DT, achieving 1.52-1.78X speed-ups. 
Figure~\ref{detail:coil},\ref{detail:timelapse} show the results on the image datasets. The detailed statistics are shown in Table~\ref{table:detail-run}. 
As is shown in the figures, PP achieves 2.4X speed-up on the Coil dataset and 5.4X speed-up on the Time-lapse dataset.

\section{Conclusion}
\label{sec:conclu}

In conclusion, we propose two parallel algorithms, multi-sweep dimension tree (MSDT) and communication-efficient pairwise perturbation (PP), to accelerate MTTKRP calculations in CP-ALS for dense tensors. These algorithms are both computationally more efficient than the standard dimension tree algorithm, and are efficient in horizontal communication. 
MSDT  reduces  the  leading  order computational  cost  by  a  factor  of $2(N-1)/N$ relative  to  the standard dimension tree algorithm. Our parallel PP algorithm reduces the communication cost to a greater extent compared to the implementations in reference~\cite{ma2018accelerating}.
Our  experimental  results  show  that  substantial performance improvements are achieved for both algorithms relative to prior approaches.
However, our theoretical analysis and results reveal that speed-ups obtained via MSDT and PP are inhibited by the lower arithmetic intensity of these two more work-efficient algorithms.

\section{Acknowledgements}
Linjian Ma and Edgar Solomonik were supported by the US NSF OAC SSI program, award No.\ 1931258.
This work used the Extreme Science and Engineering Discovery Environment (XSEDE), which is supported by National Science Foundation grant number ACI-1548562.
We used XSEDE to employ Stampede2 at the Texas Advanced Computing Center (TACC) through allocation TG-CCR180006.

\bibliographystyle{abbrv}
\bibliography{main}

\end{document}